\documentclass[12pt,a4paper]{article}
\usepackage{color,graphics,amsmath,epsfig,rotating,axodraw}

\begin{document}

\setlength{\unitlength}{1mm}


\def\ma{m_A}
\def\mhf{m_{1/2}}
\def\m0{m_0}
\def\mb{m_b}
\def\mql{m_q}
\def\mw{M_W}
\def\sw{\sin\theta_W}
\def\dMb{\Delta m_b}
\def\dMq{\Delta m_q}
\def\tb{\tan\beta}
\def\ra{\rightarrow}
\def\neuto {\tilde\chi_1^0}
\def\mneuto{m_{\tilde{\chi}_1^0}}
\def\stauo{\tilde\tau_1}
\def\staur{\tilde{\tau_R}}
\def\ser{\tilde{e_R}}
\def\smur{\tilde{\mu}_R}
\def\mt{m_t}
\def\mbmb{m_b(m_b)}
\def\mslr{m_{\tilde l_R}}
\def\wimp{$WIMP$}
\def\nn              {\notag}
\def\bce             {\begin{center}}
\def\ece             {\end{center}}

\def\mbf             {\boldmath}

\def\ti              {\tilde}

\def\a               {\alpha}
\def\b               {\beta}
\def\d               {\delta}
\def\D               {\Delta}
\def\g               {\gamma}
\def\G               {\Gamma}
\def\l               {\lambda}
\def\t               {\theta}
\def\s               {\sigma}
\def\S               {\Sigma}
\def\x               {\chi}

\def\sq              {\ti q}
\def\sqL             {\ti q_L^{}}
\def\sqR             {\ti q_R^{}}
\def\sf              {\ti f}

\def\st              {\ti t}
\def\sb              {\ti b}
\def\stau            {\ti \tau}
\def\snu             {\ti \nu}

\def\sqbar  {\Bar{\Tilde q}^{}}
\def\stbar  {{\Bar{\Tilde t}}}
\def\sbbar  {{\Bar{\Tilde b}}}

\def\Pp  {{\cal P}_{\!+}}
\def\Pm  {{\cal P}_{\!-}}

\def\ch              {\ti \x^\pm}
\def\chp             {\ti \x^+}
\def\chm             {\ti \x^-}
\def\nt              {\ti \x^0}

\newcommand{\msq}[1]   {m_{\ti q_{#1}}}
\newcommand{\msf}[1]   {m_{\ti f_{#1}}}
\newcommand{\mst}[1]   {m_{\ti t_{#1}}}
\newcommand{\msb}[1]   {m_{\ti b_{#1}}}
\newcommand{\mstau}[1] {m_{\ti\tau_{#1}}}
\newcommand{\mch}[1]   {m_{\ti \x^\pm_{#1}}}
\newcommand{\mnt}[1]   {m_{\ti \x^0_{#1}}}
\newcommand{\mq}{\mbox{$m_{\tilde{q}}$}}
\newcommand{\mhp}      {m_{H^+}}
\newcommand{\msg}      {m_{\ti g}}

\def\mw{M_\chi}
\def\sg              {{\ti g}}
\def\msg             {m_{\sg}}

\def\tW              {\t_{\scriptscriptstyle W}}
\def\tsq             {\t_{\sq}}
\def\tst             {\t_{\st}}
\def\tsb             {\t_{\sb}}
\def\tstau           {\t_{\stau}}
\def\tsf             {\t_{\sf}}
\def\sth             {\sin\t}
\def\cth             {\cos\t}
\def\cst             {\cos\t_{\st}}
\def\csb             {\cos\t_{\sb}}

\def\onehf           {{\textstyle \frac{1}{2}}}
\def\oneth           {{\textstyle \frac{1}{3}}}
\def\twoth           {{\textstyle \frac{2}{3}}}

\def\rzw             {\sqrt{2}}

\def\BR              {{\rm BR}}
\def\mev             {{\rm MeV}}
\def\gev             {{\rm GeV}}
\def\tev             {{\rm TeV}}
\def\fb              {{\rm fb}}
\def\fbi             {{\rm fb}^{-1}}

\def\over            {\overline}
\def\MSbar           {{\overline{\rm MS}}}
\def\DRbar           {{\overline{\rm DR}}}
\def\DR              {{\rm\overline{DR}}}
\def\MS              {{\rm\overline{MS}}}

\def\isajet          {{\tt ISAJET\,7.69}}
\def\softsusy        {{\tt SOFTSUSY\,1.9}}
\def\spheno          {{\tt SPHENO\,2.2.2}}
\def\suspect         {{\tt SUSPECT\,2.3}}
\def\micro           {{\tt micrOMEGAs\,1.3}}
\def\micronew        {{\tt micrOMEGAs\,2.2}}
\def\micromegas       {{\tt micrOMEGAs}}
\def\darksusy        {{\tt DarkSUSY}}
\def\calchep         {{\tt CalcHEP}}
\def\lanhep         {{\tt LanHEP}}
\def\isatools        {{\tt IsaTOOLS}}
\def\nmhdecay        {{\tt NMHDecay}}

\def\isajetnn        {{\tt ISAJET}}
\def\softsusynn      {{\tt SOFTSUSY}}
\def\sphenonn        {{\tt SPHENO}}
\def\suspectnn       {{\tt SUSPECT}}
\def\cygwin     {{\tt Cygwin}}

\def\dMb   {\Delta m_b}

\newcommand{\beqn}{\begin{eqnarray}}
\newcommand\eeqn{\end{eqnarray}}

\def\bsgamma{b\to s\gamma}
\def\bino{\tilde{B}}
\def\wino{\tilde{W}}
\def\higgsino{\tilde{H}}
\def\singlino{\tilde{S}}

\newcommand{\eq}[1]  {\mbox{(\ref{eq:#1})}}
\newcommand{\fig}[1] {Fig.~\ref{fig:#1}}
\newcommand{\Fig}[1] {Figure~\ref{fig:#1}}
\newcommand{\tab}[1] {Table~\ref{tab:#1}}
\newcommand{\Tab}[1] {Table~\ref{tab:#1}}


\bce{
 {\Large\bf  Dark matter direct detection rate in a generic
model with micrOMEGAs$\_2.2$
.} \\[10mm]

{\large   G.~B\'elanger$^1$, F.~Boudjema$^1$,  A.~Pukhov$^2$, A.~Semenov$^3$}\\[4mm]

{\it 1) Laboratoire de Physique Th\'eorique LAPTH, CNRS, Univ. de Savoie, B.P.110, F-74941 Annecy-le-Vieux Cedex, France\\

     2) Skobeltsyn Inst. of Nuclear Physics, Moscow State Univ., Moscow 119991,
Russia\\
     3) Joint Institute for Nuclear Research (JINR), 141980, Dubna, Russia  }\\[4mm]

\today}
 \ece

\begin{abstract}

We present a new module of the \micromegas~ package for the
calculation of WIMP-nuclei elastic scattering cross sections
relevant for the direct detection of dark matter  through its
interaction with nuclei in a large detector.   With this  new
module, the computation of the direct detection rate is performed
automatically for a generic model of new physics which contains a
WIMP candidate. This  model needs to be implemented within
\micronew.
\end{abstract}

\section{Introduction}

The existence of an important cold dark matter (CDM) component has been
firmly established by cosmological observations in the last few
years notably by SDSS~\cite{Tegmark:2006az} and
WMAP~\cite{Spergel:2006hy}.  A leading candidate for CDM is a new weakly interacting massive particle (WIMP).
This WIMP must be stable. Such particles arise naturally in many
extensions of the standard model ~\cite{Bertone:2004pz} from the
minimal supersymmetric standard
model~\cite{Goldberg:1983nd,Ellis:1983ew} to models of extra
dimensions~\cite{Hooper:2007qk,Cheng:2002iz,Servant:2002hb,Agashe:2004ci},
little Higgs models~\cite{Hubisz:2004ft} or models with extended
gauge~\cite{Barger:2007nv,Cvetic:1997ky} or Higgs
sectors~\cite{McDonald:1993ex,Barbieri:2006dq}. In these models
the dark matter (DM) candidate can be either a Majorana fermion, a Dirac fermion, a vector
boson or a scalar. Their masses range anywhere from a few GeV's to
a few TeV's.

Astroparticle experiments are actively pursuing searches for WIMP
DM candidates either directly through detection of elastic
scattering of the WIMP with the nuclei in a large detector or
indirectly trough detection of products of DM annihilation
(photons, positrons, neutrinos or antiprotons) in the  Galaxy or
in the Sun.

In direct detection, one measures the recoil energy deposited by
the scattering of WIMPs($\chi$)\footnote{Here we use $\chi$ to
designate the DM candidate whether a fermion, scalar or vector
boson. } with the nuclei. Generically WIMP-nuclei interactions can
be split into spin independent (scalar) and spin dependent
interations. The scalar interactions add coherently in the nucleus
so heavy nuclei offer the best sensitivity. On the other hand,
spin dependent interactions rely mainly on one unpaired nucleon
and therefore dominate over scalar interactions only for light
nuclei unless scalar interactions are themselves suppressed. In both cases, the cross-section for the WIMP nuclei
interaction  is typically low, so large detectors are required.
Many experiments involving a variety of nuclei have been set up or
are being planned. Detectors made of heavy nuclei (for example
Germanium or Xenon) currently in operation include
Edelweiss~\cite{Lemrani:2006ec}, DAMA~\cite{Bernabei:2006mx},
CDMS~\cite{Akerib:2006jk},
 Xenon~\cite{Angle:2007uj}, Zeplin~\cite{Sumner:2005wv},  Warp~\cite{Ferrari:2006pa} and KIMS~\cite{Lee:2006mz}.  Upgrades and new projects
 such as  Genius, Xmass~\cite{Kim:2006ed}, CLEAN~\cite{Nikkel:2005qj}, ArDM~\cite{Laffranchi:2007da} and Eureca~\cite{Kraus:2006pj}
 have  been proposed as well.
 Detectors made of light nuclei which are
sensitive mainly to the spin dependent interaction include Simple
~\cite{Morlat:2007pv}, Picasso~\cite{Aubin:2005rs},
Tokyo/NaF~\cite{Inoue:2003iw} and NAIAD~\cite{Alner:2005kt}. The
latter having one light (Na) and one heavy (I)  target nuclei is
actually sensitive to both spin-dependent(SD) and
spin-independent(SI) interactions. Larger versions of existing
detectors and new projects are also proposed, for example those
operating with $^3$He~\cite{Santos:2004tj,Winkelmann:2006rg}. Note
that heavy nuclei although best for probing the scalar interaction
have also a sensitivity to spin dependent interactions because of
their odd-A isotopes. Currently the sensitivity of both types of
detectors for spin dependent interactions is similar. Furthermore,
different nuclei offer a sensitivity to spin dependent
interactions on protons (for odd-proton nuclei such as $^{23}$Na,
$^{127}$I or $^{19}$F) or neutrons (for odd-neutron nuclei such as
or $^{29}$Si $^{73}$Ge,$^{129}$Xe).

 Only one experiment, DAMA,  has reported a positive
signal consistent with an annihilation cross-section $\sigma_{\chi
n}\approx 0.2-1.\times 10^{-5}$ pb for a WIMP mass around 30-100
GeV~\cite{Bernabei:2003za}. Other experiments, such as Edelweiss,
CDMS, or Xenon  have only set an upper limit on the WIMP-nucleon
annihilation cross-section\footnote{Possibilities for reconciling
DAMA results with other experiments have been considered in
Ref.~\cite{Savage:2004fn}}. The best limits were reported recently
by Xenon, $\sigma_{\chi p}^{SI}\approx 4\times 10^{-8}$ pb for a
WIMP mass around 30 GeV ~\cite{Angle:2007uj} and by CDMS,
$\sigma_{\chi p}^{SI}\approx 4.6\times 10^{-8}$ pb for a WIMP mass
around 60 GeV \cite{Ahmed:2008eu}. These values already probe a
fraction of the parameter space of the most popular CDM candidate,
the constrained minimal supersymmetric standard model
(CMSSM)~\cite{Ellis:2007ka,Trotta:2007pg} or some of its
extensions~\cite{Barger:2007nv} and poses severe constraints on a
model with a Dirac right-handed neutrino~\cite{Belanger:2007dx}.
The search for WIMPs will continue with larger detectors (around
100kg) planning to reach a level of $\sigma_{\chi n}^{SI}\approx
10^{-9}$~pb by 2010. By 2015, improved large detectors, around 1
ton, should go below the $10^{-10}$~pb level, for example Warp,
Xenon, Eureca~\cite{Kraus:2006pj} or
SuperCDMS~\cite{Akerib:2006rr}. For spin dependent interactions,
the  limits for neutrons from Zeplin~\cite{al.:2007xs} and CDMS~\cite{Akerib:2005za}
have recently been superseded by Xenon~\cite{Angle:2008we},
$\sigma^{SD}_n\approx 5.\times 10^{-3}$~pb,  while for protons the best limit from direct detection 
experiments was set by  KIMS, $\sigma^{SD}_p\approx 0.18$pb~\cite{Lee:2007qn}.
Indirect detection experiments  looking for an excess of muon neutrinos from WIMP annihilations such as Super-Kamiokande
have also set a stringent limit on
the WIMP proton cross section, $\sigma^{SD}_n \approx 3.\times 10^{-3}$~pb
~\cite{Desai:2004pq}. 
These limits are at the level or below the positive signal
reported by DAMA~\cite{Bernabei:2003za}.

The calculation of the cross-section for WIMP scattering on a
nucleon have been obtained at tree-level for different DM
candidates: neutralinos in supersymmetry(for  reviews see
~\cite{Jungman:1995df},\cite{Munoz:2003gx}), gauge bosons in UED
models~\cite{Servant:2002hb} or in little Higgs
models~\cite{Birkedal:2006fz}, right-handed neutrinos
~\cite{Agashe:2004bm} or scalars
~\cite{McDonald:1993ex,LopezHonorez:2006gr,Arina:2007tm}.
Implications of the direct detection experiments on DM models have
been explored for quite some time
~\cite{Baer:2003jb,Ellis:2005mb,Ellis:2000ds,Accomando:1999eg,
Bottino:2000jx,Kim:2002cy,Chattopadhyay:2003xi}. In the MSSM, the
most complete calculation of the neutralino nucleon scattering is
the one of Drees and Nojiri ~\cite{Drees:1993bu} that includes
higher-order effects from twist-2 operators. Public codes for DM
in the MSSM such as DarkSUSY~\cite{Gondolo:2004sc} and
Isajet~\cite{Baer:2003jb} both follow this approach for
calculating neutralino nucleon scattering. On the other hand,
\micronew, a code primarily designed for the calculation of DM
relic abundance did not, up to now, provide a module for the
computation of the direct detection rate even though an improved
tree-level computation within the MSSM has been performed for some
time~\cite{Belanger:2004ag}. This is the gap we intend to fill.
Here we describe the implementation of the direct DM detection
rate within \micronew.

Many ingredients enter the calculation of the direct detection
rate and cover both astroparticle, particle and nuclear physics
aspects. We need to know the WIMP density  and the velocity
distribution near the Earth. Since the
WIMP have small velocities, it means that the momentum transfer,
$Q^2$, is very small as compared to the masses of the WIMP and/or
nuclei. The detection rate depends of course on the WIMP nucleus
cross section. To arrive at the $\chi$-nucleus cross section one
has to first compute the interaction at the more fundamental
level, that is at the quark level. The different matrix elements
for $\chi q$ interactions that capture the dynamics of the model
in a perturbative way have to be converted into effective
couplings of WIMPs to protons and nucleons.  Finally we have to
sum the proton and neutron contribution and turn this into a cross
section at the nuclear level. The recoil spectrum of the nuclei
depends on the velocity distribution and, in view of the low
$Q^2$, is contained in the elastic form factor of the nucleus. In
this manual, we describe all these steps. Even though many of
these steps are not new, it is necessary to understand how all the
pieces are implemented in the code.

An important point to emphasize is that \micromegas, contrary to
other public codes, is not restricted to the supersymmetric model
with a neutralino DM but is applicable to a generic model of new
physics for DM~\footnote{A DM code, DM++, that can also be applied
to any new physics model has been developed by
~\cite{Birkedal:2006fz}. This code is not yet publicly available.
}. We have already in previous versions set up the code so that
any model can be implemented to give the relic density of DM, the
indirect detection rate and cross-sections relevant for collider
applications.  In the same spirit, the calculation of direct DM
detection rate in nuclei is also performed for generic models of
DM.
 More precisely the tree-level cross-section is
computed in any model and dominant QCD corrections  are taken into
account. Other higher order corrections such as the threshold
corrections to Higgs quark vertices are model dependent and are
provided only for the MSSM and its extensions (CPVMSSM, NMSSM).
The steps that go from the automatic computation of the
cross-section for WIMP scattering on quarks to a detection rate in
a large detector follow standard approaches
~\cite{Jungman:1995df}. In the spirit of the modular approach of
\micromegas,  different nuclear form factors or WIMPs velocity
distribution can easily be implemented by the user.

The paper is organised as follows, we first review the computation
of the scattering rate for DM on a point-like nucleus starting
from an effective Lagrangian for  nucleon-WIMP interactions. We
then show in Section 3 how to relate these to the  quark- WIMP
interactions and describe the method used to reconstruct the
effective Lagrangian for both scalar and spin dependent
interactions. We also describe the treatment of dominant QCD
corrections. The computation of the recoil distribution for WIMPs
scattering on nuclei taking into account nuclear form factors and
velocity distribution of WIMPs   follows in Section 4. The
functions available  in this new module are described in Section
5. Section 6 is devoted to sample results and comparisons with
other codes. The treatment of box diagrams is described in
Appendix A and  some details about nucleus spin dependent form
factors are gathered in Appendix B.

\section{Elastic scattering of WIMPs  on point-like nuclei}

The standard formalism for evaluating WIMP nuclei cross sections
was reviewed in~\cite{Jungman:1995df} with special emphasis on the
case of the neutralino DM. Here we first describe the calculation
of DM elastic scattering on point-like nuclei taking the Majorana
fermion as an example then consider different types of DM
candidates. The relation with the scattering rates on quarks,
computed automatically in our code will be presented in the next
section.

The velocity of  DM particles near the Earth should be of the same
order as  the  orbital velocity of the Sun, $v\approx 0.001c$.
Because of this small velocity, the momentum transfer is very
small as compared to the masses of the WIMP and/or nuclei. For
example for  typical  masses of WIMP, $\mw\approx 100$~GeV and of
nuclei,
 $M_A\approx 100$~GeV, the maximum transfer momentum   is
\begin{equation}
 \sqrt{-Q^2}=2v\frac{\mw M_A}{\mw+M_A}\approx 100MeV \approx 0.5
 fm^{-1}.
\label{eq:non-rel}
\end{equation}
 Thus all WIMP nucleon elastic cross sections can be calculated in
the limit of zero  momentum transfer.  The cross sections for
scattering on nuclei are  obtained from the WIMP nucleon cross
sections after folding in the nuclei form factors, these form
factors depend on the  momentum transfer.

\subsection{Scattering rate on point-like nuclei- the case of a Majorana fermion}

In the non-relativistic  limit, WIMP-nucleon elastic amplitudes
can be divided into two classes, the scalar  or spin independent
interaction and the axial-vector or spin dependent interaction.
For a spin 1/2 nucleon, interactions corresponding to multipole
will clearly vanish in the zero momentum limit. In the familiar
case of a Majorana fermion,  the effective Lagrangian
reads~\cite{Falk:1999mq}
\begin{eqnarray}
\label{neutralino_lgrgn}
  {\cal L}_{F} &=&\lambda_N\overline{\psi}_\chi\psi_\chi
  \overline{\psi}_N\psi_N
  + i\kappa_1 \overline{\psi}_\chi\psi_\chi
  \overline{\psi}_N \gamma_5 \psi_N
   +i \kappa_2\overline{\psi}_\chi \gamma_5\psi_\chi
  \overline{\psi}_N\psi_N
   + \kappa_3\overline{\psi}_\chi \gamma_5 \psi_\chi\overline{\psi}_N
   \gamma_5\psi_N\nonumber\\
   &+& \kappa_4 \overline{\psi}_\chi \gamma_\mu \gamma_5 \psi_\chi\overline{\psi}_N \gamma^\mu\psi_N
  +\xi_N\overline{\psi}_\chi \gamma_\mu \gamma_5 \psi_\chi \overline{\psi}_N \gamma^\mu \gamma_5 \psi_N
\end{eqnarray}

In the zero momentum transfer limit the operator
$\overline{\psi}\gamma_5 \psi$ vanishes while only the space
component of $\overline{\psi}\gamma_5 \gamma_\mu\psi$ and the time
component of $\overline{\psi}\gamma_\mu \psi$ remain.
 Thus the operators $\kappa_i$ are
suppressed in the limit of small momentum transfer by factors of
order $q^2/m_N^2$ and/or $q^2/m_\chi^2$ where $q^2=-Q^2$. We will ignore these
operators.  Therefore only one operator survives in each  class,
$\lambda_N$ for SI and $\xi_N$ for SD. Of course in a specific
model it is possible that the coefficient $\lambda_N$ is much
smaller than one of the coefficients $\kappa_i$,  in which case
the operator $\kappa_i$  may contribute at the same level as
$\lambda_N$ despite the $Q^2$ suppression. However in this case we
expect very small rates, much below the experimental sensitivities.

\vspace{0.5cm}
\noindent{\bf{Spin independent interactions}}\\

 For $SI$ interactions
with nucleons the effective Lagrangian  thus reads

\begin{eqnarray}
\label{neutralino_si_lgrgn}
  {\cal L}^{SI} &=&\lambda_N \overline{\psi}_\chi\psi_\chi
  \overline{\psi}_N\psi_N
\end{eqnarray}

\noindent where  $N=p,n$. The squared amplitude for a nucleon
after averaging (summing) over the polarization of incoming
(outgoing) particles is,
\begin{equation}
\label{sq_si_amplitude}
   |A^{SI}_N|^2=64\left(\lambda_{N} M_\chi M_{N}\right)^2
\end{equation}
where  $M_{N}$ is the nucleon mass. Scalar and vector WIMP-nucleon
interactions naturally induce scalar and vector WIMP-nuclei
interactions.  Summing on proton and neutron  amplitudes gives for
 WIMP-nucleus interaction at rest,
\begin{equation}
  |A^{SI}_A|^2=64M_\chi^2M_A^2(\lambda_p Z + \lambda_n(A-Z))^2
\end{equation}
where $Z$ is the nucleus charge and $A$ the total number of
nucleons. It leads to the cross section for a  WIMP scattering at
rest from a point-like nucleus
\begin{equation}
\label{SI_at_rest} \sigma_0^{SI} =\frac{4\mu_\chi^2}{\pi}\left(
\lambda_p Z + \lambda_n(A-Z)\right)^2
\end{equation}
where $\mu_\chi$ is the WIMP-nucleus reduced  mass, $\mu_\chi=\mw
M_A/(\mw+M_A)$. Note that the nucleon cross-section adds coherently
so that there is a strong enhancement for large nuclei, $\propto
A^2$ when $\lambda_p\approx \lambda_n$.

\vspace{0.5cm}
\noindent{\bf{Spin dependent interactions}}\\

The effective Lagrangian for spin dependent interactions of a
Majorana fermion at zero momentum transfer reads
\begin{eqnarray}
\label{neutralino_sd_lgrgn}
 {\cal L}^{SD} &=& \xi_N \overline{\psi}_\chi \gamma_5\gamma_{\mu} \psi_\chi
 \overline{\psi}_N\gamma_5\gamma^{\mu}\psi_N
\end{eqnarray}
It leads to the squared amplitude
\begin{equation}
|A^{SD}_N|^2=192(\xi_N  M_\chi M_N)^2 \label{sq_sd_amplitude}
\end{equation}

In order to get the  amplitudes for nuclei we have to
 sum  the spin currents produced by the protons and neutrons separately. Since we know
that the spins of protons with  the same orbital state should be
opposite,  we  expect strong compensation of currents produced by
protons as well as  those produced  by neutrons. First note that
for interactions at rest, the $\gamma_0$ component of the
pseudovector current, Eq.~\ref{neutralino_sd_lgrgn}, vanishes. The
resulting interaction $\bar{\psi}\gamma_5\gamma_i \psi$ leads to a
three dimensional vector current. This vector current  has to be
proportional to the  angular momentum  $J$. We can write for
nuclei
\begin{equation}
\label{spin_currents}
   \vec{J}_{N}^A = S_{N}^A\vec{J_A}/|J_A|
\end{equation}
where $S_N^A$ are  the expectation value of the spin content of
the nucleon N in a nucleus with A nucleons. By definition, for
protons and neutrons $S_p^p=S_n^n=0.5$ and $S_p^n=S_n^p=0$.

The second peculiarity  of the $SD$ case is a non-trivial
summation over spins. Because the matrix element is proportional
to $\vec{J}_A$ the summation over spin states in a nucleus gives a
factor
\begin{eqnarray}
\nonumber
   \sum_{s_\chi,s_\chi'} \sum_{s_A,s_A'} \sum_{1\le k,l\le3}
<s_\chi|J_\chi^k|s_\chi'><s_\chi'|J_\chi^l|s_\chi>
<s_A|J_A^k|s_A'><s_A'|J_A^l|s_A> \\
\label{J_dependence} = \sum_{1\le k,l\le3}tr(J_\chi^kJ_\chi^l)
tr(J_A^kJ_A^l)= (2J_\chi+1)J_\chi(J_\chi+1) (2J_A+1)J_A(J_A+1)/3
\end{eqnarray}
here we use $s,s'$ for labelling polarization states and $J_A$
refers to the angular momentum of a nucleus with $A$  nucleons.
After averaging over initial polarizations, a factor
$(2J_\chi+1)(2J_A+1)$ will cancel out.

Taking into account the  spin currents structure
(\ref{spin_currents}) and the $J$ dependence (\ref{J_dependence})
we can write the WIMP-nucleus squared amplitudes as
\begin{equation}
|A^{SD}|^2=256\frac{J_A+1}{J_A} \left(\xi_p S_p^A + \xi_n
S_n^A\right)^2 M_\chi^2 M_A^2 \label{sq_sd_amplitude_A}
\end{equation}
This reduces to Eq.~\ref{sq_sd_amplitude} in the special case of
the nucleon and leads to the cross section at rest for a
point-like nucleus~\cite{Jungman:1995df},
\begin{equation}
\label{sd_at_rest} \sigma_0^{SD}
=\frac{16 \mu_\chi^2}{\pi}\frac{J_A+1}{ J_A}
 \left(\xi_p S_p^A + \xi_n S_n^A\right)^2
\end{equation}

The quantities $S_N^A$ are obtained from nuclear calculations or
from  simple nuclear models, such as the odd-group model.  They
are estimated to be $\approx 0.5$ for a nuclei with  an odd number
of protons or neutrons and $\approx 0$ for an even number. Thus no
strong enhancement is expected    for SD interactions in nuclei.
The treatment of the nuclei form factors taking into account the momentum dependence will be discussed in section 5.2. 

\subsection{Generalization to other DM candidates}

To derive the formulae for elastic scattering on point-like
nuclei, we started from the effective WIMP-nucleon Lagrangian
(\ref{neutralino_si_lgrgn},\ref{neutralino_sd_lgrgn}) written for
a Majorana WIMP. In fact these can be generalized to all types of
WIMPs .
 Our aim it to give  the generic form of the
 effective Lagrangians for a fermionic, scalar and vectorial WIMP
 including the possibility of complex fields. In all cases we
 define the effective Lagrangian such that  the normalization conditions
(\ref{sq_si_amplitude},\ref{sq_sd_amplitude}) are satisfied. Here
we write only operators that contribute at $q^2=0$. As we have
argued for Majorana fermions, other operators are suppressed by
$q^2/m_{\chi(A)}^2$ and can potentially be of the same order as
the operators we consider only when both contributions to the
scattering cross section are small. Note that in the case of a
complex field, $\chi$ and $\overline{\chi}$ have in general
 different cross-sections. For each type of interaction, SI (SD)
 one can then construct two operators, one that is even  with respect to
 $\chi -\overline\chi$ interchange, $\lambda_{N,e}$($\xi_{N,e}$) and is
 the only remaining operator for Majorana's and another that is odd, $\lambda_{N,o}$
 ($\xi_{N,o}$).

For a fermion field the most general Lagrangian reads
\begin{eqnarray}
\label{fermion_si_lgrgn}
  {\cal L}_{F} &=&  \lambda_{N,e} \bar{\psi}_\chi\psi_\chi  \bar{\psi}_N\psi_N +
\lambda_{N,o} \bar{\psi}_\chi\gamma_\mu \psi_\chi  \bar{\psi}_N\gamma^\mu\psi_N\nonumber\\
 &+& \xi_{N,e} \bar{\psi}_\chi\gamma_5\gamma_{\mu} \psi_\chi  \bar{\psi}_N\gamma_5\gamma^{\mu}\psi_N
 -\frac{1}{2} \xi_{N,o} \bar{\psi}_\chi\sigma_{\mu\nu} \psi_\chi  \bar{\psi}_N \sigma^{\mu\nu}\psi_N
\end{eqnarray}
This Lagrangian leads to matrix elements which satisfy the normalization conditions
Eq.~\ref{sq_si_amplitude}(\ref{sq_sd_amplitude}) with
\begin{eqnarray}
  \lambda_N =\frac{\lambda_{N,e} \pm
  \lambda_{N,o}}{2} \;\;\; {\rm and} \;\;\; \xi_N = \frac{\xi_{N,e} \pm
  \xi_{N,o}}{2}
  \label{eq:odd-even}
\end{eqnarray}
where the $+(-)$ signs correspond to WIMP (anti-WIMP)
interactions. The special  case of a self-conjugated WIMP such as
the Majorana fermion is recovered when
$\lambda_{N,o}=\xi_{N,o}\rightarrow 0$ and
$\lambda_{N}=\lambda_{N_e}$, $\xi_{N}=\xi_{N_e}$. Note the factor
2 difference between the operator for Majorana and Dirac fermion
field, compare Eq.~\ref{neutralino_si_lgrgn} and
Eq.~\ref{fermion_si_lgrgn}. Note that the antisymmetric tensor
$\sigma_{\mu\nu}$ current effectively reduces to a vector
interaction since in the non-relativistic limit only the spatial
components contribute.

For a  scalar field only SI interactions are possible, for the
general case of a complex scalar,
\begin{eqnarray}
\label{scalar_si_lgrgn}
  {\cal L}_{S} &=& 2 \lambda_{N,e} M_\chi \phi_\chi \phi_\chi^*  \bar{\psi}_N\psi_N
  + i \lambda_{N,o} (\partial_{\mu} \phi_\chi \phi_\chi^* -  \phi_\chi
\partial_{\mu}\phi_\chi^*) \bar{\psi}_N\gamma_\mu \psi_N
\end{eqnarray}
The squared amplitude is normalized as for Majorana fermions,
(\ref{sq_si_amplitude}), with the condition (\ref{eq:odd-even})
for complex scalars. Again the case of the real scalar corresponds
to $\lambda_{N,o}=0$  and $\lambda_N=\lambda_{N,e}$.  Note that
the four-dimensional vector current $\overline{\psi}_N\gamma_\mu
\psi_N$  actually leads to a scalar interaction because only the
zeroth component of this current does not vanish in the
non-relativistic limit.

Finally for a complex vector field,
\begin{eqnarray}
\label{vector_si_lgrgn}
  {\cal L}_V &= & 2 \lambda_{N,e} M_{\chi} A_{\chi\mu} A_\chi^{\mu}
  \overline{\psi}_N\psi_N
+\lambda_{N,o} i(A_\chi^{*\alpha}\partial_\mu A_{\chi,\alpha}
-A_\chi{^\alpha}\partial_\mu A_{\chi\alpha}^*)\overline{\psi}_N\gamma_\mu\psi_N\nonumber\\
&+&  \sqrt{6}\xi_{N,e}(\partial_\alpha A^*_{\chi\beta}
A_{\chi\gamma} - A^*_{\chi\beta} \partial_\alpha
A_{\chi\gamma})\epsilon^{\alpha\beta\gamma\mu}
\overline{\psi}_N\gamma_5\gamma_{\mu}\psi_N \nonumber\\
&+& i\frac{\sqrt{3}}{2}\xi_{N,o}(A_{\chi\mu}  A_{\chi\nu}^* -
A_{\chi\mu}^* A_{\chi\nu} ) \overline{\psi}_N\sigma^{\mu\nu}\psi_N
 \end{eqnarray}
with in the special case of a real vector field
$\lambda_{N,o}=\xi_{N,o}=0$. Again the couplings are normalized as
for the fermion case both for real (\ref{sq_si_amplitude},
\ref{sq_sd_amplitude}) and complex fields (\ref{eq:odd-even}).

 In \micronew~ we assume  that DM
particles and anti-particles  have the same density. We do not
consider the case where this symmetry is broken by CP violation.
Under this assumption,  the event rate for WIMP scattering in a
large detector  is obtained after averaging over $\chi$- and
$\bar{\chi}$-nucleus cross-sections. 

\section{WIMP elastic scattering on quarks}

 The matrix elements for WIMP nucleon interactions are related to
 the more fundamental matrix elements for WIMP quarks
 interactions. These matrix elements can be easily
 written in a given model. To handle a generic model
 we rather expand WIMP quark interactions over a set of basic point-like
operators. Only a few operators are necessary in the
$q^2\rightarrow 0$  limit.

The operators that are non-zero in the non-relativistic limit are
similar to the operators introduced for nucleons in
Eq.~\ref{fermion_si_lgrgn},\ref{scalar_si_lgrgn},\ref{vector_si_lgrgn} with $\psi_N\rightarrow \psi_q$.
Those operators are listed in Table~\ref{operatorsSI} for either
spin-independent or spin-dependent interactions of a scalar
($\phi_\chi$) fermion ($\Psi_\chi$) or vector ($A_\chi^\mu$)
WIMP. Note that for the latter we use the unitary gauge.  Of
course a scalar WIMP can only contribute to spin independent
interactions. As for nucleons, the operators are separated in two
classes - odd and even, depending on the symmetry properties with
respect to quark - anti-quark exchange. For real WIMPs, odd
operators are zero by definition.
\begin{table*}[htb]
\caption{ Operators for WIMP - quark  interactions.} \vspace{.3cm}
\label{operatorsSI}
\begin{tabular}{|c|l|c|c|}
\hline
&WIMP & Even & Odd \\
&Spin&  operators      & operators   \\
\hline
&&&\\
& 0  & $2M_\chi\phi_\chi \phi_\chi^* \overline{\psi}_q\psi_q $ & $
i (\partial_{\mu} \phi_\chi \phi_\chi^* - \phi_\chi
\partial_{\mu}\phi_\chi^*)\overline{\psi}_q\gamma^\mu \psi_q $ \\
 SI &1/2 & $\overline{\psi_\chi}\psi_\chi  \overline{\psi}_q\psi_q $  & $\overline{\psi}_\chi\gamma_\mu\psi_\chi
 \overline{\psi}_q\gamma^\mu\psi_q   $ \\
& 1  & $ 2 M_{\chi} A^*_{\chi\mu} A_\chi^{\mu}
  \overline{\psi}_q\psi_q$ & +$i \lambda_{q,o}(A_\chi^{*\alpha}\partial_\mu A_{\chi,\alpha}
-A_\chi{^\alpha}\partial_\mu A_{\chi\alpha}^*)\overline{\psi}_q\gamma_\mu\psi_q$      \\
&&&\\
\hline\hline
&&&\\
&1/2 &$ \overline{\psi}_\chi\gamma_\mu\gamma_5\psi_\chi
\overline{\psi}_q\gamma_\mu\gamma_5\psi_q $
& $-\frac{1}{2}\overline{\psi}_\chi\sigma_{\mu\nu}\psi_\chi  \overline{\psi}_q\sigma^{\mu\nu}\psi_q $  \\
SD& 1  & $\sqrt{6}(\partial_\alpha A^*_{\chi\beta} A_{\chi\nu} -
A^*_{\chi\beta} \partial_\alpha A_{\chi\nu})$
  &
$i\frac{\sqrt{3}}{2}(A_{\chi\mu}  A_{\chi\nu}^* - A_{\chi\mu}^*
A_{\chi\nu} ) \overline{\psi}_q\sigma^{\mu\nu}\psi_q$
\\
&  &
$\epsilon^{\alpha\beta\nu\mu}\overline{\psi}_q\gamma_5\gamma_\mu\psi_q$  &\\
 &&&\\\hline
\end{tabular}
\end{table*}

The operators in Tables~\ref{operatorsSI} have the same
normalization as the operators for WIMP nucleon interactions, that
is the squared matrix element for $ q\chi \to q \chi $ SI interactions
 after summing/averaging over polarizations of outgoing/incoming
 particles are
\begin{equation}
\label{sq_si_amplitude2}
   |A^{SI}_{q}|^2=64\left(\frac{(\lambda_{q,e}+ \lambda_{q,o})}{2} M_\chi m_{q}\right)^2
\end{equation}
where $\lambda_{q,e}(\lambda_{q,o})$ are the coefficients of the
even (odd) operators in Table 1. In the special case of a pure
neutral WIMP, $(\lambda_{q,e}+ \lambda_{q,o})/2 \rightarrow
\lambda_{q}$. Similarly for SD interactions the normalization is
specified in Eq.~\ref{sq_sd_amplitude}.

\subsection{ Numerical approach to operator expansion}

The \micronew~ package contains  an automatic generator  of matrix
elements, CalcHEP~\cite{Pukhov:2004ca}. For each model,
\micromegas~ has a complete list of Feynman rules which specify
the model in the CalcHEP format. Therefore CalcHEP can generate
automatically amplitudes for $\chi q\rightarrow \chi q$ for any kinematics. Here
we need to generate these matrix elements at low energy in a
format that can be easily and automatically turned into  effective
operators for $\chi$-nucleon interactions. So we want to expand
automatically the Lagrangian of the model in terms of local
operators and extract the coefficients of the low energy effective
WIMP quark Lagrangian
\begin{equation}
\label{Leff}
  \hat{\mathcal L}_{eff}(x) = \sum_{q,s} \lambda_{q,s} \hat{\mathcal O}_{q,s}(x)
   +\xi_{q,s} \hat{\mathcal O}'_{q,s}(x)
\end{equation}
where $q$  is a label for quark, $s$ is a label for even and odd
operators  and $\hat{\mathcal O}_{q,s}$ ($\hat{\mathcal
O'}_{q,s}$)  are the spin independent (dependent) operators in
Table~\ref{operatorsSI}.
Traditionally the
coefficients   $\lambda_{q,s},\xi_{q,s}$ are evaluated symbolically using Fiertz identities
in the limit $q^2\ll M_A^2$. Instead in \micronew~ these
coefficients are estimated numerically using projection operators.
First we compute special matrix elements for $\chi q\rightarrow
\chi q$ scattering at zero momentum transfer. For this  we need to
add to the model Lagrangian the projection operators defined in
 Table 1. The interference between one
projection operator and the effective vertices will single out
either the spin dependent or spin independent contribution, since
the effective Lagrangian is written in an orthogonal basis.
To further separate  the coefficient of the even and odd
operators, we compute both $\chi q\rightarrow \chi q$ and $\chi
\bar{q}\rightarrow \chi \bar{q}$ matrix elements. We use the fact
that for a given projection operator the interference term in the
squared matrix elements for quarks are identical to those for
antiquarks for even operators and have opposite signs for odd
operators. Thus, taking into account the normalisation,
\begin{equation}
\label{MEplus}
 \lambda_{q,e} + \lambda_{q,o} =
\frac{-i \langle q(p_1),\chi(p_2)| \hat{S} {\hat{\mathcal
O}_{q,e}} |q(p_1),\chi(p_2) \rangle }{ \langle q(p_1),\chi(p_2)|
{\hat{\mathcal O}_{q,e}} {\hat{\mathcal O}_{q,e}}
|q(p_1),\chi(p_2)\rangle}\nonumber
\end{equation}

\begin{equation}
\label{MEminus}
 \lambda_{q,e} - \lambda_{q,o} =
\frac{-i \langle \bar{q}(p_1),\chi(p_2)| \hat{S} {\mathcal
O_{q,e}} |\bar{q}(p_1),\chi(p_2) \rangle }{ \langle
\bar{q}(p_1),\chi(p_2)| {\hat{\mathcal O}_{q,e}} {\hat{\mathcal
O}_{q,e}} |\bar{q}(p_1),\chi(p_2)\rangle}
\end{equation}
where the $S$-matrix, $\hat S=1-i {\cal L}$ is obtained from the
complete Lagrangian at the quark level.

To implement this procedure,  a new model file is created
automatically in \micronew. This file contains the model file of a
given model as well as  the auxiliary vertices of
Table~\ref{operatorsSI}. There is no need for the user to
implement these auxiliary vertices. CalcHEP then generates and
calculates symbolically all diagrams for WIMP - quark/anti-quark
elastic scattering keeping only the squared diagrams which
contains one normal vertex and one auxiliary vertex. This
corresponds to the matrix elements in
 Eq.~\ref{MEplus}.
 Note that in the file that defines the model all quarks should be
defined as massive particles.
 Vertices that depend on light quark masses  cannot be neglected, for example the couplings of Higgs to
 light quarks, since the dominant term for WIMP quark scalar
 interactions is proportional to quark masses. In particular in SUSY models
 masses of the first and second generation fermions must be included.  When converting to
   WIMP nucleon interactions this quark mass will get replaced by a nucleon
   mass so that the nucleon cross section is independent of the light
   quark masses, see Section~\ref{quark_coeff}.
Because the amplitude for WIMP scattering  on light quarks is proportionnal 
to a small quark masses, one has to be wary of numerical instabilities. 
To avoid these we consider the amplitude for
$q(p_1)\chi(p_2)\to q(p_3)\chi(p_4)$  and write the squared matrix
element is terms of the dot products $p_1.p_2$ and $p_1.p_3$.
The amplitude depends explicitly on the small momenta $p_1$ thus avoiding the numerical
instabilities in matrix elements that were found with other kinematics.

\subsection{Contribution of tensor (twist-2) operators.}

There are other point-like operators which produce the same SI and
SD amplitudes at zero momentum.  They are the operators containing
field derivatives. Here we consider the contributions of the
symmetric traceless tensor operator  to the WIMP nucleon
scattering. A complete treatment of twist-2 operators in
neutralino nucleon elastic scattering in the MSSM was first
presented in ~\cite{Drees:1993bu}. In the MSSM tensor operators
come from the momentum expansion of the denominator in the squark
exchange diagram  and are usually suppressed by a  factor
$M_N/(M_{\tilde{q}} - M_{\chi})$ with respect to the main
contribution. Thus they are expected to be mainly relevant when
the squark is not much heavier than the WIMP, for example in the
coannihilation region.

In the MSSM, the tensor operator reads \cite{Drees:1993bu},
\begin{eqnarray}
   {\cal{O}}_{q,t}&=&\frac{1}{2}(\bar{\chi}\gamma_{\mu}\partial_{\nu}\chi) {\cal{O}}_{q}^{\mu\nu}\\
{\cal{O}}_{q}^{\mu\nu} &=&\bar{q}( \gamma^\mu
\overrightarrow{\partial}^\nu
-\gamma^{\mu}\overleftarrow{\partial}^\nu +\gamma^\nu
\overrightarrow{\partial}^\mu
-\gamma^{\nu}\overleftarrow{\partial}^\mu +i m_q g^{\mu\nu} )q
\label{twist2}
\end{eqnarray}
where ${\cal{O}}_{q}^{\mu\nu}$ is a twist-2 operator (recall that
{\it twist} is defined as $dimension - spin$). Note that this
tensor operator is generically found in  other models where new
coloured particles couple to WIMP and quarks.

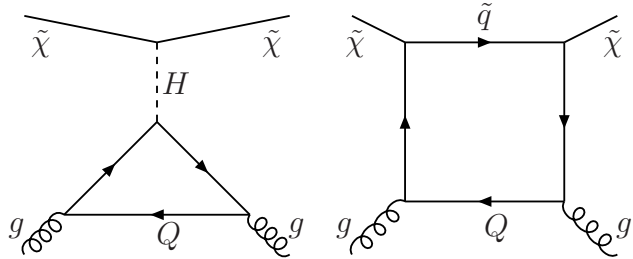
\begin{figure}
\begin{center}
{ \unitlength=1.0 pt \SetScale{1.0}
\SetWidth{0.7}      

\begin{picture}(120,100)(0,0)

\Line(0.0,90.0)(50.0,80.0) \Text(10.0,80.0)[r]{$\tilde{\chi}$}
\Line(50.0,80.0)(100.0,90.0) \Text(90.0,80.0)[l]{$\tilde{\chi}$}
\DashLine(50.0,80.0)(50.0,50.0){3.0} \Text(52.0,65.0)[l]{$H$}

\ArrowLine(15.,15.0)(50.0,50.0) \ArrowLine(50.,50.0)(85.0,15.0)
\ArrowLine(85.,15.0)(15.,15.0)  \Text(50.0,7.0)[l]{$Q$}
\Gluon(100.,0)(85.,15.){3}{3}   \Text(100.,10.)[l]{$g$}
\Gluon(0.,0)(15.,15.){3}{3}     \Text(0.,10.)[r]{$g$}

\end{picture}
\begin{picture}(120,100)(0,0)

\Line(0.0,90.0)(20.0,80.0) \Text(5.0,80.0)[r]{$\tilde{\chi}$}
\Line(80.,80.0)(100.0,90.0) \Text(95.0,80.0)[l]{$\tilde{\chi}$}

\ArrowLine(20.,80.0)(80.0,80.0) \Text(50.,85.0)[b]{$\tilde{q}$}

\ArrowLine(80.,80.0)(80.0,20.0) \ArrowLine(80.,20.0)(20.0,20.0)
\ArrowLine(20.,20.0)(20.,80.0)  \Text(50.0,10.0)[l]{$Q$}
\Gluon(100.,0)(80.,20.){3}{3}   \Text(100.,10.)[l]{$g$}
\Gluon(0.,0)(20.,20.){3}{3}      \Text(0.,10.)[r]{$g$}

\end{picture}
}\end{center} \caption{ Diagrams that contribute to WIMP-gluon
interaction via quark loops in the MSSM. } \label{QuarkLoops}
\end{figure}


To extract automatically the coefficient of the tensor operator in
the low energy WIMP quark Lagrangian we consider forward
scattering at small momentum. Indeed for such kinematics the
matrix element of the product of two scalar operators does not
depend on the collision while the product of scalar and tensor
does since

\begin{equation}
   <q(p_1),\chi(p_2)|{\cal O}_{q,t} {\cal O}_{q,e}|q(p_1),\chi(p_2)> =
-32 m_q M_{\chi} (4 (p_1.p_2)^2 -m_q^2 M_{\chi}^2)
\end{equation}
Thus the  tensor operator can be extracted by evaluating
numerically the second derivative of
\begin{equation}
         <q(p_1),\chi(p_2)| \hat{S} {\cal O}_{q,e}|q(p_1),\chi(p_2)>
\end{equation}
with  respect to $p_1.p_2$. Note that this trick does not require
that one implements the  tensor operators in the \calchep~ model.
Since the tensor operators contribute also to the amplitude at
rest, after extracting the coefficient of the tensor operator it
should be subtracted from Eq.~\ref{MEplus} to isolate the
coefficient of the scalar operator.

In typical MSSM models the correction due to an accurate treatment
of twist-2 operators is less than 1\%, in special cases though,
for example for a small mass difference between the squark and the
WIMP the contribution can be larger. The contribution of higher
spin twist-2 operators are in general further suppressed and are
not included here.
In the MSSM we have compared numerically the cross sections obtained with our 
method for extracting the effective operators, including the twist-2 operators,  
with the complete results of Drees and Nojiri~\cite{Drees:1993bu}.
For this we have implemented their analytical results, correcting a sign in the SD amplitude
between Z and squark exchange. We found  numerical agreement at the percent level.

\subsection{Nucleon form factors}
\label{quark_coeff}

In order to convert WIMP-quark amplitudes to WIMP-nucleon
amplitudes,  we need to know the values of the quark currents  inside
the  nucleon. The operators for these quark currents in a nucleon
can be extracted from experiment or estimated theoretically. In
the following we give estimates for the coefficients of the four
quark currents listed in Table~\ref{operatorsSI}

\subsubsection{Scalar coefficients}
The scalar $\bar{\psi}_q \psi_q$ current characteristic of the
SI-even interaction depends on the total number  of quarks and
anti-quarks in the nucleon. The operator $\langle
N|m_q\overline{\psi}_q \psi_q|N\rangle$  is interpreted as the
contribution of quark  $q$ to the nucleon mass, $M_N$,
\begin{equation}
\langle N| m_q\overline{\psi}_q \psi_q|N\rangle=f^N_{q}M_N
\end{equation}
where the coefficients $f^N_q$ relate nucleon and quark operators,
\begin{equation}
\lambda_{N,p}=\sum_{q=1,6} f^N_{q} \lambda_{q,p}
\end{equation}

Note there is no explicit dependence on the quark mass in the
cross section for WIMP nucleon scattering. Indeed  the quark mass
term gets transformed into a nucleon mass.
  For heavy quarks, Q, the parameter
$f^N_Q$ is induced via gluon exchange with the nucleon, see
section~\ref{sec:qcd}, and ~\cite{Shifman:1978zn}
\begin{equation}
f^N_Q= \frac{2}{27} \left( 1-\sum_{q\le 3} f^N_q \right)
\label{eq:cbt}
\end{equation}
The coefficients $f^N_{q}$ can be determined using the  value of
the light quark masses extracted from baryon masses, the ratio of
the quantities $B_q=\langle N|\bar{q}q|N\rangle$ for $u,d$ and $s$
quarks and from the value of the pion-nucleon sigma-term. It is
the latter that has the largest uncertainty, see for example
~\cite{Bottino:2001dj,Vergados:2006sy}. For light quark masses
ratio we take

\begin{equation}
\frac{m_u}{m_d}=0.553\pm 0.043, \;\;\;  \frac{m_s}{m_d}=18.9\pm
0.8
\label{eq:mumd}
\end{equation}

\noindent
We define the quantities
\begin{equation}
z=\frac{B_u-B_s}{B_d-B_s}\approx 1.49 \;\;\;\; {\rm and} \;\;\;
y=\frac{2B_s}{B_u+B_d}
\end{equation}
where $y$ denotes the fractional strange quark content of the
nucleon. $y$ is obtained from the pion-nucleon sigma term and the
quantity $\sigma_0$ which is related to the size of the SU(3)
symmetry breaking effect~\cite{Gasser:1982ap}
\begin{equation}
\label{eq:sigmaN}
\sigma_{\pi_N}=\frac{(m_u+m_d)}{2}(B_u+B_d); \;\;\; \sigma_0=
\frac{m_u+m_d}{2}(B_u+B_d-2B_s)
\end{equation}
which implies
\begin{equation} y= 1-\frac{\sigma_0}{\sigma_{\pi
N}}.
\end{equation}
Recent analyses suggest that~\cite{Pavan:2001wz}
\begin{equation}
\sigma_{\pi N}=55-73 \;{\rm MeV} \;\;\;{\rm and} \;\;\;  \sigma_0=
35\pm 5\; {\rm MeV}
\end{equation}
where $\sigma_0$ is estimated from chiral perturbation
theory~\cite{Gasser:1982ap} or from baryon mass
differences~\cite{Cheng:1988cz}.
 Defining
\begin{equation}
\alpha=\frac{B_u}{B_d}=\frac{2z-(z-1)y}{2+(z-1)y}
\end{equation}
 the parameters for light quarks in the proton write
\begin{equation}
f_d^p=\frac{2\sigma_{\pi N}}{(1+\frac{m_u}{m_d}) m_p
(1+\alpha)},\;\;\; f_u^p =\frac{m_u}{m_d} \alpha f_d^p ,\;\;\;
f_s^p=\frac{\sigma_{\pi N}y }{(1+\frac{m_u}{m_d})
m_p}\frac{m_s}{m_d}
\end{equation}
and in the neutron
\begin{eqnarray}
f_d^n=\frac{2\sigma_{\pi N}}{(1+\frac{m_u}{m_d}) m_n}
\frac{\alpha}{(1+\alpha)},\;\;\; f_u^n =\frac{m_u}{m_d}
\frac{1}{\alpha} f_d^n ,\;\;\; f_s^n=\frac{\sigma_{\pi N}y
}{(1+\frac{m_u}{m_d}) m_n}\frac{m_s}{m_d}
\end{eqnarray}

\noindent
As default values we take $\sigma_0=35$~MeV and
$\sigma_{\pi N}=55$~MeV which lead to
\begin{eqnarray}
\label{eq:scalar}
f^p_{d}=0.033, \;\; f^p_{u}=0.023, \;\; f^p_{s}=0.26 \nonumber\\
f^n_{d}=0.042, \;\; f^n_{u}=0.018, \;\; f^n_{s}=0.26
\end{eqnarray}
In \micronew, the values for these coefficients  can be changed
directly or through modification of $\sigma_0$ and $\sigma_{\pi
N}$ and the quark mass ratios. In all cases the parameters for
heavy quarks will be recalculated using Eq.~\ref{eq:cbt}. Note
that $f^N_s$ is typically larger than the value used in earlier
analyses, $f^p_S=0.118-0.14$. This is mainly due to an increase in
$\sigma_{\pi N}$ which was centered around 45 MeV
~\cite{Gasser:1990ce, Gasser:1990ap}. This large correction to
$f^N_s$ can lead to an increase by a factor 2-6 in the spin
independent cross-section for nucleons~\cite{Ellis:2005mb}.
Furthermore even with the new estimate of $\sigma_{\pi N}$, large
uncertainties remain.

\subsubsection{Vector coefficients}

The vector $\overline{\psi}_q \gamma_{\mu} \psi_q$ current in the
SI-odd interaction  is responsible for the difference between
$\chi N$ and $\overline{\chi} N$  cross sections.  The
interpretation of this current is very simple. It counts the number
of quarks minus the number of anti-quarks in the nucleon, that is
the  number of valence quarks. This current is the only one
that does  not suffer from theoretical uncertainties when
going from the WIMP- quark interaction to the WIMP-nucleon
interaction. Indeed only valence quarks contribute to the vector
current so that
\begin{equation}
\lambda_{N,o}=\sum_{q=u,d} f^N_{V_q} \lambda_{q,o}
\end{equation}
with $f_{V_u}^p=2, f_{V_d}^p=1, f_{V_u}^n=1, f_{V_d}^n=2$.

\subsubsection{Axial-vector coefficients}

The axial-vector current $\overline{\psi}_q \gamma_{\mu}
\gamma_5\psi_q$ is responsible for spin dependent interactions. It
counts the total spin of quarks and anti-quarks $q$ in the nucleon.
 Operators for axial-vector interactions in the nucleon are related to those involving
quarks,
\begin{equation}
\xi_{N,e}=\sum_{q=u,d,s} \Delta q^N \xi_{q,e}
\end{equation}
with
 \begin{eqnarray}
 2 s_\mu \Delta q^N &=& \langle N|\overline\psi_{q}\gamma_\mu\gamma_5
\psi_q|N\rangle
 \end{eqnarray}
Here $s_\mu$ is the nucleon spin and $\Delta q^N$ are extracted
from lepton-proton scattering data. The strange contribution to
the spin of the nucleon, as measured by EMC and SMC turned out to
be much larger than expected from the naive quark
model~\cite{Mallot:1999qb}. This leads to the following estimates
for the light quark contributions in the proton which have been
used in many analyses of DM spin dependent interactions,
\begin{eqnarray}
\Delta_u^p&=&0.78\pm 0.02, \;\;\; \Delta_d^p= -0.48\pm 0.02,
\;\;\; \Delta_s^p=-0.15\pm 0.02 \label{eq:spin}
\end{eqnarray}
 These early results have qualitatively been confirmed by
 HERMES~\cite{Airapetian:2007mh} and also by
 COMPASS~\cite{Ageev:2007du}.
As default values for the axial-vector coefficients we use the
latest determination of the light quark  contributions
~\cite{Airapetian:2007mh},
\begin{eqnarray} \Delta_u^p&=&0.842\pm 0.012, \;\;\; \Delta_d^p=
-0.427\pm 0.013,  \;\;\; \Delta_s^p=-0.085\pm 0.018
\label{eq:spin:compass}
\end{eqnarray}
These results are obtained in the limit of $SU(3)_F$ symmetry. It
is argued that flavour symmetry breaking effects might lead to an
additional shift in the strange quark contribution, thus making this value
compatible with the value extracted from EMC and
SMC~\cite{Mallot:1999qb}. We neglect the heavy quark contribution to the proton spin 
since they have been shown to be small~\cite{Polyakov:1998rb}. 
 The neutron quantities are simply obtained by an isospin
rotation
\begin{eqnarray}
\Delta_u^n=\Delta_d^p,\;\;\; \Delta_d^n=\Delta_u^p, \;\;\;
\Delta_s^n=\Delta_s^p
\end{eqnarray}

Note that because there can be a cancellation between the $\Delta
q$'s when summing over light quarks in the nucleon, there can be a
strong reduction in the coupling of WIMPs to neutrons or protons
when varying the quark coefficients  within the error bars.

\subsubsection{Coefficients of the $\sigma_{\mu\nu}$ term  }

 The tensor current,
$\overline{\psi}_q \sigma_{\mu\nu} \psi_q$  in SD-odd
interactions  is responsible for the difference between $\chi N$
and $\overline{\chi} N$ spin-dependent cross sections. This
current can be interpreted as the  difference between the spin of
quarks and the spin of anti-quarks in nucleons.
 Recent measurements by COMPASS~\cite{Alekseev:2007vi} and HERMES~\cite{Airapetian:2004zf}
 indicate that the antiquark contribution ($\delta \bar{u}^N+\delta \bar{d}^N$) is compatible with zero.
Forthcoming COMPASS
data on muon scattering off a proton target will provide a separate
determination of $\Delta \bar{u}^p$ and $\Delta
\bar{d}^p$~\cite{Alekseev:2007vi}. If confirmed to be zero,  it
would mean that
 the coefficients for the tensor interaction are identical
to those for the axial vector interaction. The tensor coefficients
have also been computed on the lattice, for example
Ref.~\cite{Aoki:1996pi} gives
\begin{eqnarray} \delta_u^p&=&0.84, \;\;\; \delta_d^p= -0.23,
\;\;\; \delta_s^p=-0.05. \label{eq:tensor:lattice}
\end{eqnarray}
These results were later confirmed by another lattice calculation
of the LHPC collaboration~\cite{Dolgov:2002zm}. We do not attempt
to estimate the uncertainty associated with the lattice
computation of the tensor coefficients. Note that both
collaborations also compute the axial-vector coefficients on the
lattice and they find values compatible (with larger
uncertainties) to the one obtained from COMPASS and HERMES data.
In our code it is possible to modify the input values for the
tensor coefficients independently of the axial-vector
coefficients.

\subsection{Twist-2 coefficients}

The nucleon form factors for twist-2 operators are well known in
QCD. For the operator ${\cal O}^{\mu\nu}_{q,t}$, eq.~\ref{twist2},
we have ~\cite{Drees:1993bu}
\begin{equation}
<N(p)|{\cal O}^{\mu\nu}_{q,t}|N(p)> = (p^{\mu}p^{\nu}/M_N-g^{\mu\nu}M_N  /4)
\int_0^1 (q(x)+\bar{q}(x)) x dx
\end{equation}
where $q(x)$ and $\bar{q}(x)$ are parton distribution functions
calculated at the scale $Q=M_{\tilde{q}}-M_{\chi}$. We use
the CTEQ6L distribution functions~\cite{Pumplin:2002vw}. The
computation of the box diagrams for neutralino gluon scattering in
~\cite{Drees:1993bu} leads to a loop improved twist-2 tensor form
factor containing gluon densities  instead of
 quark parton densities. However these authors argued that
the tree-level approach was more robust for $c$ and $b$ quarks
because of large logs  in the loop result,
$\log(Q/m_q)$.
 In our package we use the  tree level
formulae for both $c$ and $b$ quarks and we neglect the twist-2 form factor for $t$ quarks.
\footnote{Note  that  we evaluate the 
parton distributions  at a scale 
$Q=M_{\tilde{q}} - M_{\chi}$ rather than  $ Q=(M_{\tilde{q}}^2 - M_{\chi}^2)^{1/2}$ as suggested in~\cite{Drees:1993bu}.
In this case we find that the loop and tree-level approaches agree reasonably well for $b$-squarks when 
$M_{\tilde{b}} - M_{\chi}\approx 20$~GeV. Larger mass differences give a very small contribution from the twist-2 term
while smaller mass differences typically lead to a low value for the DM relic density.} 

\subsection{Gluon contribution and QCD corrections to the scalar amplitude}
\label{sec:qcd}

The nucleon consists of light quarks and gluons, nevertheless one
can also consider interactions of WIMPs with heavy quarks inside
the nucleon. These effectively come into play since heavy quark
loops contribute to the interactions of WIMP with gluons, see for
example the diagrams in Fig.~\ref{QuarkLoops}. It is well known
that when  Higgs exchange dominates, see for example
~\cite{Drees:1993bu},  WIMP gluon interactions via heavy quark
loops can be taken into account by considering WIMP interactions
with heavy quarks together with an estimation of the heavy quark
condensates in nucleons. Furthermore dominant QCD corrections can
also be taken into account in that case~\cite{Djouadi:2000ck}.

The anomaly of the trace of energy-momentum tensor in QCD
implies~\cite{Shifman:1978zn}
\begin{equation}
\label{Nmass} M_N\langle N|N\rangle= \langle N| \sum_{q\leq n_f}
m_q \overline{\psi}_q\psi_q(1+\gamma) +
(\frac{\beta^{n_f}}{2\alpha_s^2}) \alpha_s  G_{\mu\nu}G^{\mu\nu}
|N\rangle
\end{equation}
where $\gamma$ is the  anomalous dimension of the quark field
operator, $\alpha_s$  the strong coupling constant, $G_{\mu\nu}$
the gluon field tensor and 
$\beta^{n_f}=-\alpha_s^2/4\pi(11-2n_f/3+\alpha_s/4\pi(102-38n_f/3))$.
 In the leading order approximation for three flavours,  Eq.(\ref{Nmass}) is simplified to
\begin{equation}
\label{NmassLO} M_N\langle N|N\rangle= \langle N| \sum_{q=u,d,s}
m_q \bar{\psi}_q\psi_q -\frac{9}{8\pi} \alpha_s
G_{\mu\nu}G^{\mu\nu} |N\rangle
\end{equation}

Comparing Eq.~\ref{Nmass} for $n_f$ and $n_f+1$ one finds the
contribution of one heavy quark flavour to the nucleon mass,
relating the heavy quark content of the  nucleon to the gluon
condensate
\begin{equation}
\label{eq:heavyquark}
   \langle N|  m_Q\bar{\psi}_Q\psi_Q |N\rangle =-\frac{\Delta\beta}{2\alpha_s^2(1+\gamma)}
   \langle N|\alpha_s G_{\mu\nu}G^{\mu\nu}|N\rangle
\end{equation}
where  $\Delta\beta$ is a contribution  of one quark flavor to the
QCD $\beta$-function. This formula agrees with the effective
 $H G_{\mu\nu}G^{\mu\nu}$ vertex at small $q^2$
 \cite{Kniehl:1995tn}.
 Up to order $\alpha_s^2$,
\begin{equation}
\label{QinN} \langle N|  m_Q\bar{\psi}_Q\psi_Q |N\rangle=
-\frac{1}{12\pi}(1+\frac{11\alpha_s(m_Q)}{4\pi}) \langle N|\alpha_s
G_{\mu\nu}G^{\mu\nu}|N\rangle
\end{equation}
Keeping only the
leading order and combining with Eq.~\ref{NmassLO} will lead to
the usual relation Eq.~\ref{eq:cbt}.
Note that the  NLO terms in Eq.~\ref{Nmass} partially cancel the
effect of the NLO corrections in Eq.~\ref{QinN} so that the QCD
corrections to Eq.~\ref{eq:cbt} are small. See also
~\cite{Kryjevski:2003mh} for an alternative estimate of the heavy
quark content of the nucleon.


The simplest way to take into account dominant QCD corrections to
Higgs exchange is then to consider WIMP heavy quark interactions
through Higgs exchange and introduce an effective vertex for heavy
quarks in the nucleon with  Eq.~\ref{eq:cbt}  modified to include
one-loop QCD corrections, Eq.~\ref{QinN}. The equivalence of this
approach with  the description of the Higgs coupling to the
nucleon through gluons is confirmed by a direct computation of the
triangle diagram of Fig.\ref{QuarkLoops} in the limit where
$Q^2\ll M_Q$. Recall that the typical transfer momentum is $Q\approx
100$~MeV, Eq.~\ref{eq:non-rel}. Note that for light quarks the
corrections that would arise from their contribution to the
triangle diagram that couples a Higgs to gluon are all absorbed
into the definition of the light quark content of the nucleon.

While triangle diagrams can be treated using effective heavy quark
nucleon condensate instead of performing an explicit one-loop
calculation, such a simple treatment is not justified in general
for box diagrams.
Such an approximation would be valid only when $m_q/(M_{\tilde
q}-M_\chi)\ll 1$ as shown explicitly in ~\cite{Drees:1993bu} for the
MSSM. In that case the tree-level approach works well for c and b
quarks but would fail for t-quarks unless the associated squark is
much heavier.
  Nevertheless this is the
approximation we use by default in \micromegas, the main reason
being that in many models the contribution of the Higgs exchange
diagram is much larger than the one from the box diagrams. The
user can always ignore this simple treatment and implement a more
complete calculation of the box diagrams. For example in the case
of MSSM-like  we have implemented the one-loop computation of the
neutralino nucleon scattering of Ref.~\cite{Drees:1993bu}, see
Section~\ref{sec:susy} and Appendix A.

In a generic new physics model, new heavy coloured particles can
also contribute to the WIMP gluon amplitude, for instance squarks
in the MSSM. For heavy quarks, the computation of the triangle diagrams involving
squarks, or any other scalar colour triplet, also reduces to a
calculation of WIMP-squark scattering with  an estimation of the
squark content in the nucleon. The latter can be obtained by
calculating the contribution of squarks to the QCD
$\beta$-function just as was done for heavy quarks,
Eq.~\ref{eq:heavyquark}. Note however that the contribution of
scalars to the trace anomaly has an additional factor of 2 due to
the different dimension of scalar and fermion fields. After
substituting $\Delta \beta$ and $\gamma$ we get at order
$\alpha_s$,
\begin{equation}
\label{SQinN} \langle N|2M_{\tilde Q}^2 \phi_{\tilde
Q}^*\phi_{\tilde Q} |N\rangle= -\frac{1}{48\pi}
\left(1+\frac{25\alpha_s}{6\pi}\right) \langle N|\alpha_s
G_{\mu\nu}G^{\mu\nu}|N\rangle
\end{equation}

Thus the contribution of scalars is expected to be small because
of small scalar content in the nucleon. This relation is also
known to order $\alpha_s^2$~\cite{Djouadi:2000ck}. On the other
hand other new particles such as a heavy Majorana fermion  or a
real scalar which belong to adjoint color representation have very
large nucleon densities
\begin{equation}
\label{QinN8} \langle N|  m_Q\bar{\psi}_Q\psi_Q |N\rangle=
-\frac{1}{2\pi} \langle N|\alpha_s G_{\mu\nu}G^{\mu\nu}|N\rangle
\end{equation}

\begin{equation}
\label{SQinN8} \langle N|2M_{\tilde Q}^2 \phi_{\tilde
Q}\phi_{\tilde Q} |N\rangle= -\frac{1}{8\pi} \langle N|\alpha_s
G_{\mu\nu}G^{\mu\nu}|N\rangle
\end{equation}

In summary, in \micromegas~ we check the list of coloured
particles in the model and according to their spin  and colour
define the nucleon content for each particle using
Eq.~\ref{QinN}-\ref{SQinN8}. We then compute the contributions
from all WIMP-coloured particles processes. The coefficients of the operators for such interactions are
calculated automatically in the same manner as the coefficients for WIMP-quarks interactions
as described in Section 3.1. The case of of a color octet vector particle is not treated.

\section{The special case of supersymmetry}
\label{sec:susy}

 In the case of the MSSM, additional corrections
to both the Higgs exchange diagram and the box diagrams with
quarks or squark exchange have been computed explicitly. These
corrections can easily be extended to the case of the CPVMSSM with
complex phases and of the NMSSM with an extra singlet field.

The main contribution to the scalar neutralino nucleon cross
section involves the Higgs exchange diagram with the  Higgs
coupling to light quark pairs. While in the computation of the
relic density one could safely neglect the Higgs coupling to light
quarks, for direct detection this approximation is no longer
valid. The MSSM model file must therefore be modified accordingly.
Furthermore SUSY-QCD corrections to the Higgs exchange diagrams
can be large and should be taken into account. In particular the
gluino-squark loops can give large corrections to all
$H_i\bar{q}q$  couplings ($H_i=h,H,A$), they are also related to
the corrections to the quark masses~\cite{Djouadi:1997yw}. These
SUSY-QCD corrections will induce couplings of the quarks to the
'wrong' Higgs doublet and are enhanced by $\tan\beta$ for down
type quarks.
 The SUSY-QCD  corrections to the $H_i\rightarrow \bar{b}b$ vertices
 are already taken into account in \micromegas~ for the computation of the
relic density of DM and are also important for the computation of
$b\rightarrow s\gamma$. We generalize this to include the SUSY-QCD
corrections for all down-type quarks~\cite{Djouadi:2000ck}. As in
\micromegas$\_1.3$, we define the effective Lagrangian
\begin{eqnarray}
{\cal L}_{eff}= \sqrt{4\pi\alpha}_{QED}
\frac{\mql}{1+\dMq}\frac{1}{2 \mw\sw} \left[ - H
q\bar{q}\frac{\cos\alpha}{\cos\beta}
\left(1+\frac{\dMq \tan\alpha}{\tan\beta}\right)\right.\nonumber\\
+ i Aq\bar{q}
 \tan\beta\left(1-\frac{\dMq}{\tan\beta^2}\right)
+h q\bar{q}\left. \frac{1}{\cos\beta} \left
(1-\frac{\dMq}{\tan\alpha\tan\beta}\right)\right]
 \end{eqnarray}
where $q=d,s,b$. We use the same conventions as in ~\cite{Belanger:2004yn}
 for the supersymmetric model parameters and our code is compatible with the  
 SUSY Les Houches Accord (SLHA)~\cite{Skands:2003cj}.
For light quarks, $\Delta m_{d,s}$ includes only
the squark gluino loop  while additional electroweak corrections
which are important for the third generation are included in
$\Delta m_b$~\cite{Belanger:2004yn}.
 These corrections are much smaller and
can be neglected for up type quarks. As we mentioned before there
is no explicit dependence on the quark mass for the scalar
neutralino nucleon cross section, nevertheless the effect of
SUSY-QCD corrections remains.

Other vertices that contribute to neutralino quark interactions
and also have a  dependence on the light quark mass are the
$\tilde\chi_1^0 \tilde{q} q$ vertices. These vertices involve a
coupling of the Higgsino component which is proportional to the
quark mass as well as a coupling of the gaugino component which
depends on the squark mixing matrix. Since the  off-diagonal
element of the 2X2 mixing matrix for a given squark flavour is
also proportional to the quark mass, mixing cannot be neglected
even for the first two generations. The MSSM model therefore needs
to be modified to add these extra terms in vertices for light
quarks.  In general, input parameters for \micromegas~ are taken
from the SLHA~\cite{Skands:2003cj} and
assume no mixing for the first two generations, we define the
mixing angle as

\begin{equation}
\tan 2\phi_q=- 2 \frac{m_q \tilde{A}_q}{M^2_{\tilde q_R}-
M^2_{\tilde q_L}}
\end{equation}
where $\tilde{A}_u=A_u-\mu/\tan\beta$ for up-type quarks and
$\tilde{A}_d=A_d-\mu \tan\beta$ for down-type quarks, $m_q$ is the
quark mass including SUSY-QCD corrections~\cite{Djouadi:1997yw}.
 For third generation
sfermions, the mixing angle is taken directly from the spectrum
calculator as given in the SLHA~\cite{Skands:2003cj}\footnote{In
the case of Isajet, we set $A_d=A_s=A_b$ and $A_u=A_c=A_t$.  Note
that this assumption is not strictly correct but the mixing in the
down-squark sector is in any case dominated by the $\mu\tan\beta$
term while in the up-squark sector it is mostly relevant for the
top squark.}.

The importance of QCD and $\Delta m_b$ corrections in a few sample
supersymmetric models is displayed In Table~\ref{tab:qcddmb}. A
more detailed description of the parameters of the  models
considered can be found in Tables~\ref{tab:MSSM},~\ref{tab:other}.
The QCD corrections in Eqs.~\ref{QinN},\ref{SQinN} are relevant
only for heavy quarks and will always lead to an increase of the
spin independent  cross section, typically around $5\%$. The
SUSY-QCD corrections on the other hand  depend on the sign of
$\mu$ and are enhanced  at large values of $\tan\beta$. In the
decoupling limit, these corrections affect mainly the couplings of
heavy Higgses as well as the couplings of squarks while the light
Higgs, which often gives the dominant contribution to the scalar
cross section remains unchanged. For $\mu>0$, $\Delta m_q>0$ and
the cross section decreases, so that the net effect of the QCD and
SUSY-QCD corrections remains small. For $\mu<0$, $\Delta m_q<0$
and both the heavy Higgs and squark contribution increases. In the
sample model KP of Table~\ref{tab:qcddmb} the SUSY QCD corrections
lead to a mild decrease of $\sigma^{SI}_n$, this is because there
is a destructive interference between the dominant light Higgs
exchange and the heavy Higgs and squark diagrams.

\begin{table*}[h]
\caption{Effect of higher-order corrections on $\sigma_{n}^{SI}$
in sample SUSY models.} \label{tab:qcddmb} \vspace{0.3cm}
\begin{center}
\begin{tabular}{|l|l|l|l|l|l|}
\hline
          & BP & KP & IP  & NUH &MSSM1\\ \hline
$\tb$     & 10  & 40 & 35 & 30 & 10 \\\hline
 $\mu$    & + & - & + & +  & +\\ \hline\hline
$\Omega h^2$  & 0.101 & 0.101 & 0.113 &0.088 &0.100   \\\hline
$\sigma_{n}^{SI}\times 10^{9}pb$   &  &  &  &  &  \\
tree-level       &  8.25 & 8.63 & 26.1 &  9.12& 19.1\\\hline
 QCD             &  8.50 & 9.13 & 26.8 &  9.36& 19.8\\\hline
$\Delta m_b$     &  7.53 & 8.52 & 20.0 &  7.75& 18.1\\\hline 
QCD+$\Delta m_b$ &  7.77 & 9.02 & 20.5 &7.97& 18.7\\\hline
 QCD+$\Delta m_b$ +box  &  7.78 & 9.02 & 20.5 & 7.97 & 18.7\\\hline
\end{tabular}
\end{center}
\end{table*}

Finally in the MSSM and its extensions we also compute more
precisely one-loop corrections. In that case we replace the
tree-level treatment of heavy quarks by a one-loop computation of
the process $\chi g\rightarrow \chi g$ including triangle and box
diagrams involving quarks and squarks.  The corrections to the
dominant Higgs exchange triangle diagram discussed above are
always included so the bulk of the corrections are taken into
account with either option. The one-loop correction from the box
diagram is obtained by modifying the denominators entering the
tree-level diagrams, details can be found in Appendix A. This
method reproduces the complete results of  Drees and
Nojiri~\cite{Drees:1993bu}. We have implemented these analytic
results and provide a special routine that can be used for
comparing the two approaches.
 In the last two lines of Table~\ref{tab:qcddmb} we  compare the results using heavy quark
currents  with the ones including  the more complete treatment of
loop corrections (box diagrams). In mSUGRA models where the Higgs
exchange diagrams dominate, discrepancies are below the per-cent
level. Larger discrepancies can be found in the general MSSM,
where, based on explicit calculations,  we expect loop corrections
to be of order $m_q/(M_{\tilde q}-M_\chi)$. This can be large for
heavy quarks specially when squark masses are of the same order as
the neutralino masses.

\section{Scattering rates on nuclei: form factors and velocity distribution}

To get the rate for direct detection of WIMPs as a function of the
recoil energy of the nucleus we must take into account both the
finite velocity of WIMPs and the nucleus form factor.

\subsection{Spin independent interactions}
First consider the simplest case of the $SI$ interaction. We note
that ignoring form factor effects, we expect isotropic scattering
in the center on mass frame. This means that in the laboratory
frame for a WIMP  of velocity $v$  one gets a constant
distribution over the recoil energy in the interval
$0<E<E_{max}(v)$. In the  non-relativistic approximation,
\begin{equation}
    E_{max}(v)=2\left( \frac{v^2 \mu_\chi^2 }{M_A}\right)
\end{equation}
For an incoming WIMP with a fixed velocity the recoil energy
distribution is thus of the form
\begin{equation}
\label{dSQdE_vixed_v}
    \frac{d\sigma^{SI}_A}{dE}=\sigma^{SI}_0 \frac{\Theta(E_{max}(v)-E)F_A^2(q)}{E_{max}(v)}
\end{equation}
where $F_A(q)$ is the nucleus form factor which  depends on the
transfer momentum $q=\sqrt{2 E M_A}$.

DM particles have a certain velocity distribution, $f(v)$. After
integration over incoming velocities, the distribution of the
number of events over the recoil energy reads
\begin{eqnarray}
\label{eq:dNdE:SI}
\frac{dN^{SI}}{dE}=\frac{2M_{det}t}{\pi}\frac{\rho_{0}}{M_\chi}
F_A^2(q)\left(\lambda_p Z + \lambda_n(A-Z) \right)^2I(E)
\end{eqnarray}
where $\rho_0$  is the DM density near the Earth, $M_{det}$ the
mass of the detector and $t$ the exposure time and
\begin{eqnarray}
I(E)&=&\int_{v_{min}(E)}^{\infty} \frac{f(v)}{v} dv\\
v_{min}(E)&=&\left(\frac{E M_A}{2 \mu^2_\chi}\right)^{1/2}
\end{eqnarray}

For $SI$ interactions,   the form factor is a Fourier transform of
the nucleus distribution function,
\begin{equation}
\label{Fourier}
    F_A(q)=\int e^{-iqx} \rho_A(x) d^3x
\end{equation}
where  $\rho_A(x)$ is normalized such  that $F_A(0)=1$. In our
package, we use the Fermi distribution function
\begin{equation}
     \rho_A(r)=\frac{c_{norm}}{1+exp((r-R_A)/a)}
\end{equation}
where $c_{norm}$ is fixed by the normalization condition. The
resulting form factor is often referred to as the Woods-Saxon form
factor~\cite{Jungman:1995df}.  The parameters $R_A$ and $a$ can be
extracted from muon scattering data. Compilations of these
parameters  for several  nuclei are available
~\cite{Fricke,deVries}. A two-parameter fit to these tables
~\cite{Lewin:1995rx} gives
\begin{equation}
R_A= 1.23 A^\frac{1}{3}-0.6 {\rm fm}  \label{eq:RA}
\end{equation}
for a surface thickness, $a=0.52$~fm. This is the  default value
we use for all nuclei. We have tested this interpolation.
Using the data on the first three radial moments of charge distribution~\cite{Fricke}
we calculate the form factor at a recoil energy of 15~keV, a value slightly above
threshold for DM detectors. We then calculate  the half-density radius
$R_A$ which reproduces this value.
Fig.~\ref{fig:nuclearff} shows that for various nuclei, the value
of $R_A$ extracted this way is well approximated by
Eq.~\ref{eq:RA}.

\subsection{Spin dependent interactions}

The $SD$ case  features the same kinematics as the case just
discussed. The simplest approximation would  be to  assume that
the $S_N^A$ coefficients have the same q-dependence as the $SI$
form factor (\ref{Fourier}). For a more precise evaluation of the
$q$ dependence,  one should note that when $q\ne 0$  the vectors
$\vec{J_p}$ and $\vec{J_n}$ are not  collinear anymore. As a
result, three form factors need to be introduced. They  correspond
to the three coefficient of the quadratic function of
$\xi_p^2,\xi_n^2$ and $\xi_p\xi^n$ in the squared amplitude,
Eq.~\ref{sd_at_rest}. Equivalently, one can construct the
isoscalar and isovector combinations
\begin{eqnarray}
a_0=\xi_p +\xi_n\\
a_1=\xi_p -\xi_n
\end{eqnarray}
so that the SD recoil energy distribution for a fixed WIMP
velocity reads

\begin{eqnarray}
\label{dSsddE_vixed_v}
    \frac{d\sigma^{SD}_A}{dE}=\frac{16 \mu^2_\chi}{2J_A+1}
    (S_{00}(q)a_0^2+S_{01}(q)a_0a_1+S_{11}(q)a_1^2)
    \frac{\Theta(E_{max}(v)-E)}{E_{max}(v)}.
\end{eqnarray}
The coefficients $S_{00}(q)$, $S_{11}(q)$, and $S_{00}(q)$   are
the nuclear structure functions which take into account both the
magnitude of the spin in the nucleon and the spatial distribution
of the spin. They are normalized such that \cite{Bednyakov:2006ux}
\begin{eqnarray}
\nonumber
S_{00}(0)=C(J_A)(S_p+S_n)^2\\
\label{Sxx_0}
S_{11}(0)=C(J_A)(S_p-S_n)^2\\
\nonumber
S_{01}(0)=2C(J_A)(S_p+S_n)(S_p-S_n)\\
\nonumber {\rm where}\;\; C(J_A)=\frac{(2J_A+1)(J_A+1)}{4\pi J_A}
\label{S00}
\end{eqnarray}
With this normalization, one recovers  the cross section at rest,
Eq.~\ref{sd_at_rest}, starting from Eq.~\ref{dSsddE_vixed_v}.

After taking into account the velocity distribution, the
distribution for the number of events over the nuclei recoil
energy for spin dependent interactions reads
\begin{eqnarray}
\label{eq:dNdE:SD}
\frac{dN^{SD}}{dE}=\frac{8M_{det}t}{2J_A+1}\frac{\rho_{0}}{M_\chi}
(S_{00}(q)a_0^2+S_{01}(q)a_0a_1+S_{11}(q)a_1^2)I(E)
\end{eqnarray}

\begin{figure}[ht]
  \centerline{\epsfig{file=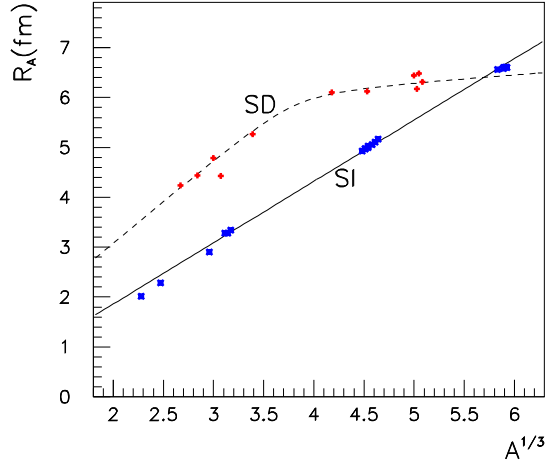, width=9cm}}
  \caption{$R_A$ for SI (full/blue) and SD (dotted/red) Gauss distribution form factors for various
  nuclei,  the curves correspond to the  approximations~\ref{eq:RA} and ~\ref{eq:rasd}.}
\label{fig:nuclearff}
\end{figure}

The form factors are calculated from detailed nuclear models
including the momentum dependence. The list of SD form factors
implemented in \micromegas~  is given in the Appendix. For nuclei
for which the form factor has not been computed precisely, we can
describe the SD form factors  with a Gauss distribution,
\begin{equation}
\label{Gauss}
   S_{ij}(q)=S_{ij}(0)exp(-q^2 R^2/4)\;\;,
\end{equation}
where $S_{ij}(0)$ is defined in Eq.~\ref{S00} and $S_p$ and $S_n$ can be found in ~\cite{Jungman:1995df,Bednyakov:2004xq}.
 To find out
the effective nucleus radius  to be used in the Gauss distribution, we perform a
fit to  the known form factors. 
First we observe that with a recoil energy $E=15$~keV the form factors have a  similar  $q$ dependence  (within 10\%), 
\begin{equation}
   \frac{ S_{11}(q)}{S_{11}(0)}\approx \frac{S_{00}(q)}{S_{00}(0)}\approx \frac{S_{01}(q)}{S_{01}(0)}.
\end{equation} 
Thus, we can  use the same effective radius for all form factors. 
To extract the $A$ dependence of $R$ we assume that 
Z exchange dominates, which means that $S_{11}$ is
the dominant form factor. We thus obtain

\begin{equation}
    R_{A}= 1.7 A^{1/3}-0.28 -0.78\big( A^{1/3}-3.8 + \sqrt{(A^{1/3}-3.8)^2+0.2}\;\;\big)\;\;fm
\label{eq:rasd}
\end{equation}
A comparison between this approximate formula and the value
extracted from the form factors tabulated in \micromegas~ shows
that the approximation works quite well, see
Fig.~\ref{fig:nuclearff}. For light nuclei ($A<100$), $R_A=1.5
A^{1/3}$~fm is also a very good approximate formula
~\cite{Ellis:1991ef}. Some dependence  on the DM model remains
because of the different $q$ behaviour of $S_{ij}$. This
dependence  has been shown to be small ~\cite{Ressell:1997kx}.

\begin{table*}[htb]
 \caption{Number of events per kg/day $\times 10^5$ for $5-50$~keV recoil energy for SD
interactions on various nuclei for model BP. Comparison between  two sets of form factors  
 and the Gauss approximation in eq.~\ref{Gauss}  }\label{tab:sdff}
\begin{center}
\begin{tabular}{|l|l|l|l|l|l|l|l|l|l|l|l|}
  \hline
&${}^{19}F$&${}^{23}Na$&${}^{27}Al$&${}^{29}Si$&${}^{39}K$&${}^{73}Ge$&${}^{93}Nb$&${}^{125}Te $&${}^{127}I $&${}^{129}Xe $&${}^{131}Xe$ 
\\\hline
$Sxx$ Tab.8  &  16.3 &2.03  &3.31  &2.32   &2.67    &4.57  &2.44  &4.60     &0.79     &5.97   & 1.55 \\ \hline
Gauss        &  16.4  &2.05  &3.36  &2.26   &2.70    &4.67  &2.50  &4.68     &0.75     &6.25   & 1.46\\ \hline \hline
$Sxx$ Tab.9  &       &2.13  &      &2.32   &        &7.92  &      &5.54     &1.27     &4.37   & 1.37 \\ \hline
Gauss         &       &2.02  &      &2.19   &        &7.81  &      &6.06     &1.23     &4.60   & 1.30\\ \hline 
\end{tabular}
\end{center}
\end{table*}

We compute  the number of events for SD interactions for the
nuclei listed in Table~\ref{tab:sdff} in the region $5<E<50$~keV.
For this we choose the sample model BP in Table~\ref{tab:MSSM} and
compare the results using the precise form
factorsof Table 8 and Table 9~\cite{Bednyakov:2006ux}  with the ones obtained by using a
Gauss distribution with  the approximated formula
Eq.~\ref{eq:rasd} as well as the coefficients $S_p$ and $S_n$ in
Eq.~\ref{S00}. We find that the approximate results with the Gauss distribution agree well with the exact form factors,
the agreement is usually better than between different sets of form factors.  
Even for small nuclei where the interpolation for
the radius $R_A$ does not work as well there is a very good
agreement between the different sets of  form factors, see
Table~\ref{tab:sdff}. This is because for light nuclei the
momentum is very small hence the form factor does not need to be
known precisely. Note that the Gauss approximation would not work as well at large recoil energies where the form factor decreases too rapidly.

\subsection{Velocity distribution of dark matter}

The nuclear recoil energy measured in direct detection experiments
depends on the WIMP velocity distribution in the rest frame of the
detector (\ref{eq:dNdE:SI},\ref{eq:dNdE:SD}). This in turn
depends on the WIMP velocity distribution in the rest frame of the
galaxy and the Earth velocity with respect to this frame. The
latter is determined from various observations. The measurements
of the velocity of Sun and others objects close to the Sun give a
value both for the velocity of rotation of the LSR(local standard
of rest) \cite{Kerr:1986hz}
\begin{equation}
\label{v0}
    v_0=220\pm 20 km/s
\end{equation}
and for the peculiar velocity of Sun in this system
\cite{Dehnen:1997cq, Galactic_Astronomy} \footnote{Here we use the
galactic co-ordinates (X,Y,Z) where X is toward the Galactic
center, Y in the direction of rotation and Z toward the north
Galactic pole.}
\begin{equation}
\label{vpec}
    \vec{v}_{pec}= (10.0, 5.2, 7.2) km/s
\end{equation}
The Earth velocity with respect to the galactic frame is thus the
sum  of $\vec{v}_0=(0,v_0,0)$, $\vec{v}_{pec}$ and of the Earth
velocity in the solar system.  Assuming that the Earth's orbit is
circular and that the axis of the ecliptic lies in the Y-Z plane,
the Earth velocity in Galactic co-ordinates is
\begin{equation}
\label{ve}
    \vec{v}_e =v_e \big( -\sin(2\pi t),\;\;\sin\gamma \cos(2\pi t),
\;\;\cos\gamma\cos(2\pi t) \big)
\end{equation}
where  $v_e=29.79km/s$, $\gamma=30.5^\circ$ and $t$ is in years.
More precise expressions, taking into account an elliptic orbit
and the orientation of the axis of the ecliptic can be found in
\cite{Green:2003yh}.

The velocity distribution of DM particles on the Earth is obtained
from the DM velocity distribution in the rest frame of the Galaxy,
$F_{GRF}$,
\begin{equation}
\label{fv}
   f(v) = \int \delta(v-|\vec{V}|) F_{GRF}(\vec{V}-\vec{v}_0-\vec{v}_{pec}-\vec{v}_e) d^3\vec{V}
\end{equation}
Because the mass of Galaxy is finite  there is some $v_{max}$ such
that $F_{GRF}=0$ for  $|V|>v_{max}$, astronomical
observations\cite{Smith:2006ym}
 give  the 90\% confidence interval
$$
   498 km/s < v_{max} < 608 km/s
$$
with  a median likelihood of $v_{max} = 544 km/s$.

There are several models of DM velocity
distribution\cite{Belli:2002yt}, they are correlated with the DM
density distribution. The simplest and most widely used model to
describe the DM density is the isothermal sphere
model~\cite{Jungman:1995df}. In such a model the DM velocity
distribution corresponds to  a Maxwellian distribution. In our
package we have implemented a truncated Maxwellian distribution,

\begin{equation}
\label{maxwell}
      F_{GRF}(\vec{V}) \sim exp( -|\vec{V}|^2/{\Delta}V^2)\Theta(v_{max} -|\vec{V}|)
\end{equation}
which leads to
\begin{equation}
\label{fv_max}
  f(v)=c_{\rm{norm}}\left[
exp\left(-\frac{(v-v_1)^2}{{\Delta}V^2}\right) -
exp\left(-\frac{min(v+v_1,v_{max})^2}{{\Delta}V^2}\right)\right]
\end{equation}
where $\Delta V=v_0$ in the isothermal model, $c_{\rm{norm}}$ is
fixed by the normalization condition
$$
  \int^\infty_0 f(v)dv =1
$$
and
$$
v_1=|\vec{v}_0+\vec{v}_{pec}+\vec{v}_e|\approx v_0+(v_{pec})_y +
v_e \sin\gamma \cos(2\pi t)
$$
Note that the Earth motion around the  Sun leads to a 7\%
modulation effect of $v_1$ and in turn to a modulation of the
signal in direct detection experiments. This  modulation was
investigated in \cite{Green:2003yh,Belli:2002yt} for a large set
of models  of DM spatial distributions. The DM velocity
distribution close to the Sun could be quite different from the
Maxwell distribution. For example condensation of cold DM in
clumps and streams will lead to a delta-function distribution.
This function has also been implemented in the code.  The impact
of the different velocity distributions on the limit on the
detection rate of neutralinos was also discussed in
~\cite{Bottino:2005qj}. In order to allow for deviations from the
isothermal model,  in the implementation of the Maxwellian
distribution in \micromegas~ we treat ${\Delta}V$, $\rho_0$, $v_1$
as free parameters. Furthermore alternative models for DM
distribution can always be implemented by the user.

The DM density near the SUN is estimated to be in the range
$\rho_0=0.1-0.7$~GeV/cm$^3$~\cite{Yao:2006px}. As default, we take
the commonly used value $\rho_0=0.3$~GeV/cm$^3$.

\section{Description of routines}

All routines are available both in C and Fortran and in both cases
they have the same names and the same arguments. For simplicity,
here we describe only C routines. The types of parameters are
listed in \\
sources/micromegas.h for C routines and
 in sources/micromegas\verb|_|f.h for Fortran routines.

\noindent
$\bullet$ \verb|setProtonFF(scalar, ps_vector, sigma)|\\
$\bullet$ \verb|setNeutronFF(scalar, ps_vector, sigma)|

These  two routines allow to redefine the default values of the
coefficients describing the quark contents of nucleons, see
Section~\ref{quark_coeff}. Each parameter is an array of type
$double$ of dimension 3 whose elements correspond to the form
factors for $d$, $u$, and $s$ quarks respectively. Default values
of form factors are displayed in Table {\ref{NFF}}. To keep the
default value for one set of coefficients, use NULL (NOFF in
Fortran).

\begin{table*}[hb]
\caption{Default values of quark form factors in proton and
neutron.} \label{NFF} \vspace{0.3cm}
\begin{center}
\begin{tabular}{|l|l|l|l||l|l|l|}
\hline
  &\multicolumn{3}{|c||}{proton} & \multicolumn{3}{|c|}{neutron} \\ \cline{2-7}
current & d & u & s& d & u & s\\ \hline scalar
&0.033&0.023&0.26&0.042&0.018&0.26  \\ \hline
$\gamma_5\gamma_{\mu}$& -0.427& 0.842&-0.085&0.842 &-0.427& -0.085  \\
\hline $\sigma_{\mu\nu}$&-0.23& 0.84  & -0.046 & 0.84 &-0.23  &  -0.046  \\
\hline
\end{tabular}
\end{center}
\end{table*}

\noindent $\bullet$
\verb|getScalarFF(|$m_u/m_d,m_s/m_d,\sigma_{\pi
N},\sigma_0,$\verb|FF_proton,FF_neutron )|\\
 This routine computes the scalar coefficients for quark contents in the
nucleon from the mass ratios $m_u/m_d,m_s/m_d$ as well as from
 $\sigma_{\pi N}$ and $\sigma_0$ following
 Eq.~\ref{eq:mumd}-\ref{eq:sigmaN}. It can be used to find the input  values
 of the quark coefficients  in  \verb|setProtonFF| and
 \verb|setNeutronFF|. $\sigma_{\pi N}$ and $\sigma_0$ are specified in MeV.
 The coefficients are stored in  \verb|FF_proton| and in
 \verb|FF_neutron|,
both arrays of dimension 3 and of type $double$.

\noindent
$\bullet$ \verb|FeScLoop(sgn, mq,msq,mne)|\\
This function computes the  loop K-factor for MSSM-like models, Eq.~\ref{eq:kfactor} with a
spin 1/2 WIMP and a scalar "squark". It allows to replace
the denominators for  $s,u$-channel diagrams as specified in eq.~\ref{eq:propagator}. Here
\verb|mq, msq,mne| refer to  quark, squark, and WIMP  masses.

\noindent
 $\bullet$
\verb|nucleonAmplitudes(qBOX,pAsi,pAsd,nAsi,nAsd)|\\
This routine calculates amplitudes for WIMP-nucleon elastic
scattering at zero momentum. \verb|pAsi(nAsi)| are spin
independent amplitudes for protons(neutrons), whereas
\verb|pAsd(nAsd)| are the corresponding spin dependent amplitudes.
Each of these four parameters is an array of type $double$ and
dimension 2. The first (zeroth) element of these arrays gives the
$\chi$-nucleon amplitudes whereas the second element gives
$\overline{\chi}$-nucleon amplitudes. Amplitudes are normalized
such that the total cross section for either $\chi$ or $\overline
\chi$ cross sections
\begin{equation}
\sigma_{tot}=\frac{4\mu_\chi^2}{\pi}(|A^{SI}|^2+3|A^{SD}|^2)
\label{eq:norm}
\end{equation}

The qBOX parameter specifies
a function like \verb|FeScLoop()| that is used to improve the tree-level calculation
in order to reproduce the results of the box-diagram calculation. The function \verb|FeScLoop|
included in our package can be used for spin 1/2 WIMPS and scalar "squarks".
To obtain the tree-level result, substitute  $qBOX=NULL$ ($qBOX=NoLoop$) in C(Fortran).
\verb|nucleonAmplitudes| returns a value different from zero only
when there is an internal problem in calculation.

\noindent
$\bullet$  \verb|MSSMDDtest(loop, &pS,&pV,&nA,&nV)|\\
This routine computes   the  proton(neutron) scalar, $pS$($nS$)
and vector, $pV$,($nV$) one-loop or tree-level amplitudes in the
MSSM using directly the formulae of Drees and Nojiri
~\cite{Drees:1993bu}. If $loop=0$ then calculations are done at
tree level assuming heavy quark condensates in the nucleon,
otherwise the one-loop results for neutralino interactions with
gluons are used.  The QCD and SUSY-QCD corrections are included.
Amplitudes are normalized according to (~\ref{eq:norm}). This
routine exists only in the C-version.

\noindent
$\bullet$ \verb|fDvMaxwell(v)| \\
returns  $f(v)/v$ where $f(v)$ is defined in Eq.~(\ref{fv_max}).
The argument $v$ is expressed in $km/s$. This function is used as
argument of the \verb|nucleusRecoil| function described below.

\noindent
$\bullet$ \verb|SetfMaxwell(DV,v1,vmax)|\\
sets parameters  for \verb|fDvMaxwell|. The arguments correspond
to the parameters ${\Delta}V$, $v_1$, and $v_{max}$ in
Eq.~(\ref{fv_max}) in $km/s$ units.  Unless this routine is called
the program uses the default values (220,225.2,700).

\noindent $\bullet$ \verb|SetfDelts(v)|\\
sets velocity of DM for $\delta$-function distribution.

\noindent $\bullet$ \verb|fDvDelts(v)|\\
indicates to nucleusRecoil that the velocity distribution is a
delta function which is non-zero for the parameter specified in
\verb|SetfDelts(v)|.

\noindent $\bullet$ \verb|SetFermi(C,B,a)| sets parameters for
A-dependence of the Fermi half radius,
    $R_A= CA^{1/3}+B$
and for the surface thickness, \verb|a|. Default values are given in
Eq.~\ref{eq:RA}.

\noindent
$\bullet$ \verb|nucleusRecoil(rho,fDv,A,Z,J,S00,S01,S11,qBOX,dNdE)|\\
This is the main routine of the  direct detection module. The
input parameters \verb|rho|,  \verb|fDv|, specify the DM velocity
distribution while $A$, $Z$, $J$, $S00$, $S01$, $S11$ specify
properties of detector material. The return value gives the number
of events per day and per kilogram of detector material. The
distribution over recoil energy is stored in the array $dNdE$.
This array has to be of type $double$ with 200 elements. The value
in the $i^{th}$ element corresponds to
$$
\frac{dN}{dE}|_{E=i*keV}
$$
in units of ${\rm (1/keV/kg/day)}$. For a complex WIMP,
\verb|nucleusRecoil| averages over WIMP and
$\overline{\rm{WIMP}}$.

The input parameters are:\\
\verb|rho| - density of DM near the Earth in $GeV/cm^3$;\\
\verb|fDv| - this parameter is a function which specifies the DM
velocity distribution, \verb|fDv|$=f(v)/v$ where $f(v)$ is defined
in Eq.(\ref{fv}), the velocity $v$ is given in $km/s$ and
\verb|fDv(v)| in $(s/km)^2$. The function
\verb|fDvMaxwell| described above is the default function that can be used here.\\
\verb|A| - Atomic number of nucleus;\\
\verb|Z| - Number of protons in the nucleus;\\
\verb|J| - nucleus spin.\\
\verb|qBOX| - a parameter needed by \verb|nucleonAmplitudes|, see the description above.\\
$S00(p)$, $S01(p)$, $S11(p)$ are nucleus form factors for
spin-dependent interactions. They  are functions of the momentum
transfer in $fm^{-1}$ (argument of  $double$ type). These form
factors are assumed to be normalized as in Eq.~\ref{S00}.

Nucleus form factors for a wide set of nuclei are implemented in
\micromegas. The available form factors are listed in
Tables~\ref{SDFF},~\ref{SDFFA} in  Appendix B  and are extracted
from the review article of Bednyakov and
Simkovic~\cite{Bednyakov:2006ux}. These functions are named
\begin{equation}
\label{SxxFunc}
   S\{xx\}\{Nucleus\; Name\}\{Atomic\; Number\}[A]
\end{equation}
where $xx$ is either $00$,$11$ or $01$. The last character is
optional and is used to distinguish different implementations of
form factor for the same isotope.

For convenience our package has predefined constants for nucleus
charge
$$
   Z\_\{Name\}
$$
and nucleus spin
$$ J\_\{Name\}\{atomic\_number\} $$
for all isotopes listed in Tab.\ref{SDFF},\ref{SDFFA}.

\noindent
For example, a call to \verb|nucleusRecoil| for
${}^{73}Ge$ detector should be
\begin{verbatim}
N=nucleusRecoil(0.3,fDvMaxwell,73,Z_Ge,J_Ge73,S00Ge73,S01Ge73,S11Ge73,FeScLoop,dNdE);
\end{verbatim}

\noindent
$\bullet$ \verb|int PlotSS(Sxx,A, title, Emax)|\\
allows a visual check of the spin dependent form factors. This
routine opens a new window where   \verb|Sxx| is displayed as
function of the recoil energy. Here \verb|Sxx| stands for  any of
functions in  Tab.~\ref{SDFF},~\ref{SDFFA}, $A$ is the atomic
number, \verb|title| defines the text that will be displayed on
the screen and \verb|Emax| is the upper limit of the recoil energy
for the plot.

The \verb|PlotSS| routine is based on CalcHEP~\cite{Pukhov:2004ca}
plot facilities. By a click of the  mouse on one point in the plot
area one gets information about the value of X/Y coordinates for
that point. To close the window press the $'Esc'$ key. Pressing
some other key opens the menu which displays \verb|min| and
\verb|max| values of the function and allows to change limits of
the $Y$ axis and switch between linear/logarithmic scale of $Y$
axis. For additional help press the \verb|F1| key.

\noindent
$\bullet$\verb|nucleusRecoil0(0.3,fDv,A,Z,J,Sp,Sn,qBOX,dNdE)|\\
is similar to the  function \verb|nucleusRecoil| except that it
uses Eq.~\ref{Sxx_0} to define $S00(0)$, $S11(0)$, $S01(0)$ and
the q-dependence of these form factors is the one associated  with
the Fermi distribution  with radius, $R_A$ given in
Eq.~\ref{eq:rasd}. This function can for example be used for light
nuclei, such as $^3He$, or for any other nuclei where the more
precise form factor has not been included in Tab. ~\ref{SDFF}. For
all nuclei listed in Tab.~\ref{SDFF} as well as for  ${}^{1}H$
($S_p=0.5,S_n=0$) and ${}^{3}He$\cite{Engel:1989ix}, form factors
have been predefined with names
\begin{equation}
\label{SpSnConst}
    Sp\{Nucleus\; Name\}\{Atomic \;Number\}\;\;\; {\rm and} \;\;\;   Sn\{Nucleus\; Name\}\{Atomic\; Number\}
\end{equation}
 These constants correspond to the form
factors listed in  Tab.\ref{SDFF} extracted from the review
article~\cite{Bednyakov:2004xq}.  They all satisfy Eq.\ref{Sxx_0}.
Table ~\ref{tab:sdff} shows how well the approximation works.

Two auxiliary routines are provided to work with the energy
spectrum computed with
\verb|nucleusRecoil| and  \verb|nucleusRecoil0|.\\
$\bullet$\verb|displayRecoilPlot(dNdE,title,E1,E2)|\\
plots the  generated energy distribution dNdE. Here \verb|title|
is a character string specifying the title of the plot and
\verb|E1,E2| are minimal and maximal values for the displayed
energy. \verb|E1,E2| have to be an integer value in keV units,
\verb|E2| has to be smaller than 200. This  routine  has the same
service facilities as \verb|PlotSS| described above. Note that the
plots can be saved in Latex format as well as in GNUPLOT or PAW
formats.

\noindent
$\bullet$\verb|cutRecoilResult(dNdE, E1, E2)|\\
calculates the number of events in an energy interval defined by
values $E1$,$E2$ in keV units.

\section{Sample results}

\begin{table*}[htb]
\vspace{0.3cm} \caption{SI and SD cross sections  in sample SUGRA
models.} \label{tab:MSSM} \vspace{0.3cm}
\begin{tabular}{|l|l|l|l|l|l|l|l|l|l|}
\hline

          & AP & BP & CP& DP & IP & KP & MP &  NUG& NUH\\ \hline
 $m_0$    & 130 &70 & 90 & 120 & 180 & 2500 &1100 & 1620 &250   \\ \hline
 $M_{1/2}$& 600 &250 & 400 & 500 & 350 & 550 &1100 & 300 &530   \\ \hline
 $A_0$    &0 &-300 & 0 & -400 & 0 & -80 & 0 & 0 & 0   \\ \hline
$\tb$     &  5 & 10 & 10 & 10 & 35 & 40 & 50 & 10 & 30 \\\hline
 $\mu$ & + & + & + & - & + & - & + & + & +   \\ \hline\hline
Masses &  &  &  &  &  &  &  &  &    \\ \hline

$\neuto$       &  248.4 & 97.9  & 161.7 & 207.4 & 141.3 & 223.5 & 476.3 & 109.4 & 218.6\\\hline 
$\chi_1^+$     &  466.8 & 183.0 & 302.6 & 395.7 & 264.4 & 284.1 & 893.3 & 156.7 & 404.3\\\hline
$\tilde{l}_1$  &  257.1 & 107.6 & 170.5 & 213.8 & 152.3 & 2129. & 816.7 & 1607. & 254.4\\\hline 
$\tilde{t}_1$  &  952.9 & 362.5 & 649.9 & 774.6 & 581.8 & 1730. & 1836. & 999.3 & 848.2\\\hline
           $h$ &  111.5 & 111.0 & 112.5 & 114.2 & 112.4 & 118.6 & 119.2 & 113.7 & 115.3\\\hline
           $A$ &  888.2 & 420.5 & 576.4 & 769.0 & 426.2 & 1547.6 & 927.4 & 1614.& 497.8 \\\hline\hline

$\Omega h^2$ & 0.126 & 0.101& 0.115 & 0.107 & 0.113 & 0.101 & 0.128 & 0.119 &0.088   \\
\hline $\sigma_{p}^{SI}\times 10^{9}$pb &  &  &  &  &  &  &  &  &
\\
set A & 0.98 & 7.50 & 2.48 & 0.019 & 19.6 & 9.03 & 0.509 & 46.9  & 7.65\\
set B& 1.82  & 15.9 & 4.98 & 0.055 & 44.1 & 14.4  & 1.06 & 85.7 & 16.7 \\
set C& 0.77  & 5.49 & 1.87 & 0.011 & 13.9 & 7.67 & 0.375 & 37.2  & 5.50\\\hline
$\sigma_{p}^{SD}\times 10^{6}$pb &  &  &  &  &  &  &  &  &    \\
set A' & 0.301 & 5.00 & 1.97 & 0.360 & 3.78& 140.  & 0.096 & 510.  & 2.22    \\
set B' & 0.230 & 3.69 & 1.56 & 0.261 & 3.05& 129.  & 0.082 & 470.  & 1.89   \\\hline
$\sigma_{n}^{SD}\times 10^{6}$pb &  &  &  &  &  &  &  &  &    \\
set A' & 0.251 & 4.34 & 1.55 & 0.322 & 2.89& 87.0  & 0.069 & 317.  & 1.58    \\
set B' & 0.334 & 5.89 & 2.01 & 0.443 & 3.69& 99.1  & 0.085  & 362.  & 1.93
\\\hline
N $\times 10^3$/kg/day &  &  &  &  &  &  &  &  &    \\
 $^{73}Ge$& 0.19 & 3.00  & 0.70 & 0.0071 & 6.08& 2.41  & 0.056 & 20.3  & 1.68    \\
$^{131}Xe$ & 0.31 & 5.35 & 1.18 & 0.0085 & 10.5& 3.36 & 0.089  & 31.6  & 2.79   \\
 \hline
\end{tabular}
\end{table*}

\begin{figure}[ht] \vspace{-1cm} 
\centerline{\epsfig{file=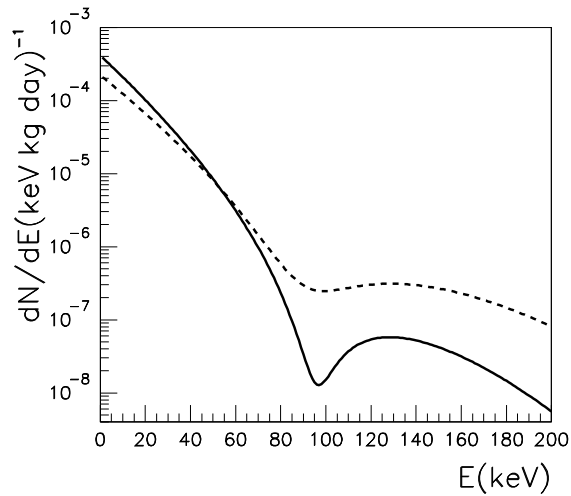, width=9cm}}
 \vspace{-1cm} 
  \caption{dN/dE for a $^{131}Xe$ detector in two mSUGRA models
specified in Table~\ref{tab:MSSM}, BP (full) and KP (dash).}
\label{fig:display}
\end{figure}

The predictions for the spin independent  cross sections on
protons and spin dependent cross sections on neutrons and protons
are listed in Table~\ref{tab:MSSM} in the case of the MSSM with
input parameters fixed at the GUT scale. The first seven models
(AP to MP) are inspired by the mSUGRA benchmarks of
Ref.~\cite{Battaglia:2003ab}. The input parameters  have been
modified such that the prediction for $\Omega h^2$ falls near the
central WMAP value when the top quark mass is fixed to the value
measured at Tevatron, $m_t=171.4$~GeV~\cite{Brubaker:2006xn}.
 The parameters of  Model BP were adjusted to the ones of the SPA1A benchmark~\cite{AguilarSaavedra:2005pw}.
 The spectrum calculator used is SuSPECT~\cite{Djouadi:2002ze}.  The last two models are
non-universal SUGRA models. In model NUG, the gaugino masses are
non-universal at the GUT scale with $M_3=0.7 M_2=0.7 M_1$ leading
to a lighter coloured sector than in the universal case. In model
NUH only the Higgs masses are non-universal with $M_{H_u}=2 m_0$
and $M_{H_d}=-0.6 m_0$. For each model the predictions for the SI
cross section are given for three sets of values for the
coefficients describing the quark density contents in the nucleon.
Set A correspond to the default values, Eq.~\ref{eq:scalar}, while
the input parameters for set  B ($\sigma_{\pi
N}=70,\sigma_0=35$~MeV) and set C ($\sigma_{\pi
N}=55,\sigma_0=40$~MeV) are varied within the expected range,
Eq.~\ref{eq:sigmaN}. Note that these do not correspond to the most
extreme choices, yet the predictions for the cross sections can
vary by a factor of 2 or 3. For the SD cross sections, set A' and
set B' correspond respectively to the old (Eq.~\ref{eq:spin}) and
new (Eq.~\ref{eq:spin:compass}) estimates of the quark density
coefficients. In Table~\ref{tab:MSSM} the predictions for the
number of events per day and per kg of detector material for
$^{73}Ge$ and $^{131}Xe$ are also presented. Here we assume the
default values for the quark density coefficients as well as for
the velocity distribution. The distribution of the number of
events as function of the energy is also computed. The results for
models BP and KP  for a detector made of $^{131}Xe$ are displayed
in Fig.~\ref{fig:display}.

We have also compared our results with other public codes. For
this we have used  the same spectrum calculator (here we have
taken Isajet$\_7.75$)  and have removed from \micromegas~ the QCD
and SUSY QCD corrections which are neglected in other codes. For a
given set of quark coefficients, our results for $\sigma^{SI}$ are
in very good agreement with Isajet$\_7.75$ for the dominant
contribution due to Higgs exchange, differences appear in the
squark exchange contribution. For SUGRA models this can lead to
25\% differences between the two codes. For $\sigma^{SD}$, our
prediction is usually below the one of Isajet, differences between
the two codes can reach a factor 4. This can be traced back to a
sign discrepancy between the Z and the squark exchange diagram
\footnote{The direct detection module is being improved in
Isajet, we expect much better agreement between the two codes in
the next public version, Isajet$\_7.76$~\cite{AB_private}.}. As
concerns DarkSUSY$\_4.1$, we have discrepancies that can reach
25\% for $\sigma^{SI}$, this is due to the Higgs exchange diagram.
In DarkSUSY  the running mass is used in the $hqq$ vertex while
the pole mass is used when estimating the quark scalar
coefficient, whereas in \micromegas~ we use the same mass
everywhere as we have discussed in Section~3.3. Our predictions
for $\sigma^{SD}$ are in very good agreement with those of
DarkSUSY for  the Z exchange contribution. We however have a
factor 2 difference in the squark exchange diagram which can lead
to 50\% discrepancies in $\sigma^{SD}$ for our SUGRA test models.
Increasing the squark exchange contribution by a factor 2 in
DarkSUSY, we recover very good agreement with \micromegas~\footnote{The DarkSUSY code has been updated and is now in good agreement with \micromegas~\cite{PG_private}}.

\begin{table*}[htb]
\begin{center}
\vspace{0.3cm} \caption{Comparison with Isajet$\_7.75$ and
DarkSUSY$\_4.1$ in sample mSUGRA models} \label{tab:ds_Isa}
\vspace{0.3cm}
\begin{tabular}{|l|l|l|l|l|l|l|}
\hline
          & AP & BP & CP& DP & IP &  MP \\ \hline
 $m_0$    & 130 &70 & 90 & 120 & 180 &1100    \\ \hline
 $M_{1/2}$& 600 &250 & 400 & 500 & 350 & 1100 \\ \hline
 $A_0$    &0 &-300 & 0 & -400 & 0 &  0  \\ \hline
$\tan\beta$     &  5 & 10 & 10 & 10 & 35 &  50 \\\hline
 $\mu$ & + & + & + & - & + & +    \\ \hline\hline

$\sigma_{p}^{SI}\times 10^{9}$pb &  &  &  &  &  &
\\
micrOMEGAs & 0.466 & 3.65 & 1.17 & 0.0067 & 9.57 & 0.16  \\
Isajet& 0.448 & 2.85 & 1.01 & 0.0025 & 7.18 & 0.14  \\
Isajet'& 0.460 & 3.64 & 1.16 & 0.0067 & 9.45 & 0.16  \\
Darksusy& 0.357  & 2.89 & 0.895 & 0.0054 & 7.54 & 0.118  \\\hline
$\sigma_{p}^{SD}\times 10^{6}$pb &  &  &  &  &  &    \\
micrOMEGAs & 0.248 & 4.44 & 1.66 & 0.306 & 3.19 & 0.068    \\
Isajet& 0.87 & 16.7 & 4.72 &1.37  &8.16 & 0.141  \\
Isajet'& 0.241 & 4.31 & 1.62 & 0.297 &3.11 & 0.067  \\
DarkSUSY & 0.370 & 6.79 & 2.29 & 0.506 &4.21 & 0.082  \\
DarkSUSY' & 0.252 & 4.49 & 1.68 & 0.315 &3.19 & 0.067  \\\hline

$\sigma_{n}^{SD}\times 10^{6}$pb &  &  &  &  &  &    \\
micrOMEGAs & 0.203 & 3.75 & 1.29 & 0.267 & 2.41& 0.0489     \\
Isajet& 0.45 &8.49  &2.49  & 0.694 & 4.35 & 0.077   \\
Isajet'& 0.198&3.66  & 1.26  &0.260  &2.36  &  0.0478  \\
DarkSUSY & 0.254 &4.71  &1.54  & 0.353 &2.80 & 0.054   \\
DarkSUSY' & 0.201 &3.70  &1.27  & 0.266 &2.36 & 0.047   \\\hline
\end{tabular}
\end{center}
\end{table*}

\begin{table*}[htb]
\begin{center}
\caption{SI and SD cross sections on protons in non-SUGRA models,
all masses in GeV.} \label{tab:other}
\begin{tabular}{|l|l|l|l|l|}
\hline
           & MSSM1 & LH & RHN1 & RHN2 \\ \hline
           &  $M_1=200$    & $f=1000$ & $g_Z=0.002$ & $g_Z=0.0066$   \\
           &  $M_2=400$  & $M_H=220$ & $g_H=0.025$ & $g_H=0.025$   \\
           &  $\mu=350$ & $\kappa=1$ & $M_{Z'}=5000$ & $M_{Z'}=5000$ \\
           &  $m_{\tilde{q}_R}=300$   & $\kappa_1=0.5$ & $M_{\nu_R}=46.$ & $M_{\nu_R}=900.$   \\
           &  $m_{\tilde{l}_R}=200$ & $\sin\alpha=\frac{1}{\sqrt{2}}$ & $M_H=200.$ &  $M_H=200.$ \\
           &  $\tb=10$ &  &  & \\\hline
           Masses (GeV) & &  &  &   \\\hline
 $\chi$         & 193.2 & 150.2 & 46. & 900. \\\hline 
$\tilde{l_1}$  & 204.7 & 701.7 & 5000.& 5000.\\\hline 
$\tilde{t_1}$  & 326.6& 991.9 & 5000. & 5000. \\\hline
$h$            & 115.6 & 220.0 & 200. & 200. \\\hline\hline
$\Omega h^2$   & 0.100 & 0.109& 0.100 & 0.151   \\
\hline $\sigma_{n}^{SI}\times 10^{9}pb$ & &    &  &
\\
 A &   18.9 & 0.33 &  47.9   & 584. \\
 B &   38.7  & 0.60 &  46.2   & 578. \\
 C &   14.2  & 0.27 &  48.4   & 586. \\\hline
$\sigma_{p}^{SD}\times 10^{6}pb$ &  &    &  &\\
 A'& 0.47 & $3.1\times 10^{-6}$ &  0.317   & 3.59 \\
 B'& 0.039 & $6.5\times 10^{-6}$ & 0.293   & 3.31    \\\hline
$\sigma_{n}^{SD}\times 10^{6}pb$ &&   &  &\\
A' & 3.17 & $2.7\times 10^{-8}$ & 0.197    & 2.22  \\
B' & 7.21 & $3.7\times 10^{-7}$ & 0.224    &  2.53   \\\hline
\end{tabular}
\end{center}
\end{table*}

In the sample models of Table~\ref{tab:other}  the dependence on
the axial vector coefficients for the spin dependent cross-section
is far less important (within 30\% between sets A' and B') than
for the scalar cross section. Furthermore much smaller variations
are observed if one just sticks to  the uncertainty associated
with the latest experimental results, Eq.~\ref{eq:spin:compass}.
This is specific to SUGRA models where one finds that the Z
exchange diagram completely dominates over the squark exchange
diagram. This is valid in all models where Z exchange dominates.
In the more general MSSM  with weak scale input parameters, it is
possible to find models where the squark exchange diagrams
contribute significantly to the spin dependent cross section and
can furthermore interfere with the Z exchange diagram. This can
lead to a strong reduction of either the proton or neutron cross
section as shown for a sample MSSM model in Table~\ref{tab:other}
where  the most relevant parameters are specified in the Table, in
addition   we set $m_{\tilde{q}_L}=M_A=-A_t=1$~TeV,
$m_{\tilde{l}_L}=500$~GeV and $M_3=800$~GeV. Finally sample
results for the case of a gauge boson in the Little Higgs model
(LH){~\footnote{We use the version of the Little Higgs model
implemented into CalcHEP and available at http://hep.pa.msu.edu,
more details can be found in~\cite{Belyaev:2006jh}.}} and of a
Dirac right-handed neutrino (RHN1 and RHN2) are listed in
 Table~\ref{tab:other}. Note that in
this table the values for SI cross sections on neutrons only are
given, in most models these numbers are similar to the ones of
protons except in the RH neutrino model where the cross section on
protons is very much suppressed. In this model the small
dependence on the quark coefficients is due to the Higgs exchange
diagram. The results for the RH neutrino model agree with the ones
of ~\cite{Belanger:2007dx}.

\section{Installation}

The package can be obtained from the web page
\verb|wwwlapp.in2p3.fr/lapth/micromegas|. All instructions for
installing the package can be found in ~\cite{Belanger:2006is}.
Contrary to previous versions, the user must download only one
file  \verb|micromegas_2.2.tgz| which contains the full
implementation of different models such as the MSSM,
CPV-MSSM~\cite{Belanger:2006qa},
NMSSM~\cite{Hugonie:2007vd},\cite{Belanger:2005kh} as well as the
little Higgs model (LHM)~\cite{Belyaev:2006jh} and the
right-handed neutrino model(RHNM)~\cite{Belanger:2007dx}.
Facilities to incorporate other new models remain as in
micrOMEGAs$\_{}2.0$, specific instructions on how to install new
models are given in ~\cite{Belanger:2006is}. Note that the identifiers \verb|S0| and \verb|V5|
are now reserved for the auxiliary fields used to generate automatically the operators in
Table~\ref{operatorsSI}, they
cannot be used to designate particles in the model. Note also
that since micrOMEGAs searches for quarks in the list of particles by identifying the PDG codes,
care has to be taken to implement these correctly in the model file.
The direct
detection module is not compatible with earlier versions of the
supersymmetric models, since light quark masses and the corresponding
couplings of light quarks to Higgses need to be introduced. To run the code we provide
for each model one sample file, \verb|main.c| (\verb|main.F| in
Fortran). This sample program can be used to compute  the relic
density of DM, the cross sections for direct and indirect
detection, the cross sections at colliders and decay widths.
Various options can be set in that program depending on the need
of the user, the various switches available are commented in the
main file. The sample programs cycle2.c and cycle5.c found in the MSSM directory
will reproduce the numerical results in
Tables~\ref{tab:qcddmb},\ref{tab:MSSM}.

\section{Summary}

The new module for computation of the cross-section for WIMP
scattering on nucleus in \micromegas~ described here applies to
any type of CDM candidate, whether Majorana or Dirac fermion,
scalar or vector boson. After  the new  model is implemented
within \micromegas, the mass, the spin and the interactions of the
WIMP candidate are computed automatically. Because the nucleon has
an important light quarks component, WIMP interactions with light
quarks in the nuclei often dominate over those of heavy quarks.
For scalar interactions the amplitude for WIMP quark scattering is
proportional to the quark mass, thus the mass of light quarks have
to be taken into account and incorporated into the model file. In
the specific case of the MSSM this means expanding the model file
since  for the relic density calculation all fermions of the first
two generations were taken to be massless. Furthermore within this
model, higher-order corrections to the $H\bar{q}q$ vertices for
down-type quarks such as SUSY-QCD corrections are taken into
account. For all models, we use an effective vertex for WIMP
interactions with heavy quark in the nuclei, this takes into
account dominant QCD corrections to Higgs exchange diagrams. A
more precise treatment is available for MSSM-like models.

 We have also explicitly  shown on a few examples the impact of
 varying the coefficients for the quark density content of the
 nucleon. Uncertainties are very large for scalar interactions and
 are in general more under control for spin dependent interactions
 that arise through Z exchange. For other contributions to spin
 dependent interactions the uncertainties can be quite large
 although the value predicted is often several orders of magnitude
 below the present limit.

\section{Acknowledgements}
We thank A. Djouadi and E. Pilon for clarifications on QCD
corrections, V. Shevchenko for discussions on the tensor quark
coefficients, A. Bednyakov for discussions on the nuclear form
factors and A. Belyaev for  providing the CalCHEP code for the
little Higgs model and for discussions on the \micromegas-Isajet
comparison.  We also thank S. Kraml for testing various parts of
our code.  Part of this work was performed during the CERN Theory
Institute on "LHC-Cosmology Interplay", A.P. and G.B. thank the
organisers for their warm hospitality. This work was supported in
part by the GDRI-ACPP of CNRS and by the French ANR project
ToolsDmColl, BLAN07-2-194882. The work of A.P. was supported by the 
Russian foundation for Basic Research, grants RFBR-08-02-00856-a, RFBR-08-02-92499-a.

 \section*{Note added}  
An online tool for computation of direct detection rates with \micronew~ 
has been set up by Rachid Lemrani, see \\
\verb|http://pisrv0.pit.physik.uni-tuebingen.de/darkmatter/micromegas_g/|.

\renewcommand{\theequation}{A-\arabic{equation}} 
  \setcounter{equation}{0}  
  \section*{Appendix A - Box diagrams}  

As shown explicitly by Dress and Nojiri, the tree-level
calculation overestimates s-channel amplitudes when
$M_{\tilde{q}}-M_\chi<m_q$. In that case one should instead
compute completely the box diagrams or use the procedure described
here.
 In the MSSM, the Lagrangian for neutralino-quark-squark interaction has the generic form
\begin{equation}
   {\cal{L}}_Y = \bar{q}(a + b\gamma_5)\chi\tilde{q}+h.c.
\end{equation}
and leads to the WIMP-quark scattering amplitude at tree-level
\begin{equation}
 A= \frac{1}{4}\big(  \frac{b^2}{M_{\tilde{q}}^2-(M_{\chi}+m_q)^2}
-  \frac{a^2}{M_{\tilde{q}}^2-(M_{\chi}-m_q)^2}\big)
\label{eq:box_tree}
\end{equation}
The explicit computation of the box diagram~\cite{Drees:1993bu} leads to an amplitude of the form
\begin{equation}
A_{box}= \frac{1}{4}\left((b^2-a^2)F_D(m_q,M_{\tilde{q}},M_{\chi})
+ (a^2+b^2)F_S(m_q,M_{\tilde{q}},M_{\chi})\right)
\end{equation}
where $F_D$ and $F_S$ are loop functions. A simple modification of the propagators in
\ref{eq:box_tree} will therefore reproduce the amplitude $A_{box}$,
\begin{equation}
   \frac{1}{M_{\tilde{q}}^2-(M_{\chi}\pm m_q)^2} \to K(\pm 1,m_q,M_{\tilde{q}},M_{\chi})
\label{eq:propagator}
\end{equation}
where
\begin{eqnarray}
K(s,m_q,M_{\tilde{q}},M_{\chi}) &=&
F_D(m_q,M_{\tilde{q}},M_{\chi})+s F_S(m_q,M_{\tilde{q}},M_{\chi})\nonumber\\
 &=&\frac{3}{2} m_q \big(m_q I_1 -\frac{2}{3}M_{\chi}^2 I_3-s M_{\chi}( I_2 -\frac{1}{3} I_5 -\frac{2}{3}
 M_{\chi}^2I_4)\big)
\label{eq:kfactor}
\end{eqnarray}
and
\begin{eqnarray}
I_1&=&\int_0^1 dx \frac{x^2-2x+2/3}{D^2}\\
I_2&=&\int_0^1 dx \frac{x(x^2-2x+2/3)}{D^2}\\
I_3&=&\int_0^1 dx \frac{x^2(1-x)^2}{D^3}\\
I_4&=&\int_0^1 dx \frac{x^3(1-x)^2}{D^3}\\
I_5&=&\int_0^1 dx\frac{x(1-x)(2-x)}{D^2}\\
D&=& x^2 M_\chi^2 +x(M_{\tilde{q}}^2 -m_q^2-M_\chi^2)+m_q^2
\end{eqnarray}
The  modification of  the  denominators of  tree level diagrams in
micrOMEGAs is achieved by  a small modification of the CalcHEP
code. Obviously this trick will work in any MSSM-like model where
the WIMP is a fermion that interacts with quarks  via scalar
quarks,
 in particular in the  NMSSM and the CPV-MSSM.
The same trick can be generalized to  other models, for this
however one has to compute the appropriate loop factors which
depend on the WIMP and 'squark' spin. Note that the box diagrams
for WIMP scattering on gluons includes also diagrams where gluons
couple to squarks. When the WIMP scattering amplitude is computed
with the loop K-factor option, \micromegas~ omits  the tree-level
processes involving squarks (or in general colored triplet that
are not quarks) assuming that their contributions are included in
the K-factor.

\renewcommand{\theequation}{B-\arabic{equation}}
  \setcounter{equation}{0}  
  \section*{Appendix B - Spin dependent nucleus form factors }  

\begin{table*}[h]
\caption{ Nucleus SD form factors realized in \micromegas }
\label{SDFF} \vspace{0.3cm}
\begin{tabular}{|l|l|l|l|l|}
\hline
Identifier& Isotope &  Ref& data in \cite{Bednyakov:2006ux}& comments  \\
\hline
SxxF19   & ${}^{19}F$   &\cite{Divari:2000dc}                  & Eq. 7  & \\
SxxSi29  & ${}^{29}Si$  &\cite{Divari:2000dc}                  & Eq. 14 & \\
SxxNa23  & ${}^{23}Na$  &\cite{Ressell:1997kx} & Eq. 9       &      \\
SxxTe125 & ${}^{125}Te$ &\cite{Ressell:1997kx} & Eq. 18, Tab. IV  & Bonn-A potential\\
SxxI127  & ${}^{127}I$  &\cite{Ressell:1997kx} & Eq. 20  & Bonn-A potential\\
SxxXe129 & ${}^{129}Xe$ &\cite{Ressell:1997kx} & Eq. 21 Tab. IX       & Bonn-A potential\\
SxxXe131 & ${}^{131}Xe$ &\cite{Ressell:1997kx} & Eq. 21 Tab. IX       & Bonn-A potential\\
SxxAl27  & ${}^{27}Al$  &\cite{Engel:1995gw}                  & Eq. 11       & \\
SxxK39   & ${}^{39}K$   &\cite{Engel:1995gw}                  & Eq. 15       & \\
SxxGe73  & ${}^{73}Ge$  &\cite{Dimitrov:1994gc}                     & Eq. 17       & \\
SxxNb92  & ${}^{93}Nb$  &\cite{Engel:1992qb}                     & Tab. II      & Scanned  plot \\
\hline
\end{tabular}
\end{table*}

\begin{table*}[htb]
\caption{ Alternative nucleus  SD form factors.} \label{SDFFA}
\vspace{0.3cm}
\begin{tabular}{|l|l|l|l|l|}
\hline
Identifier& Isotope &  Ref& data in \cite{Bednyakov:2006ux}& comments  \\
\hline
SxxSi29A & ${}^{29}Si$  &\cite{Ressell:1993qm}                 & Eq. 12 & \\
SxxNa23A & ${}^{23}Na$  &\cite{Divari:2000dc}                  & Eq. 10 & \\
SxxTe125A& ${}^{125}Te$ &\cite{Ressell:1997kx} & Eq. 18, Tab.IV  &  Nijmegen II potential\\
SxxI127A & ${}^{127}I$  &\cite{Ressell:1997kx} & Eq. 19, Tab.VI  &  Nijmegen II potential\\
SxxXe129A& ${}^{129}Xe$ &\cite{Ressell:1997kx} & Eq. 21 Tab. IX    &  Nijmegen II potential\\
SxxXe131A& ${}^{131}Xe$ &\cite{Ressell:1997kx} & Eq. 21 Tab. IX    &  Nijmegen II potential\\
SxxGe73A & ${}^{73}Ge$  &\cite{Ressell:1993qm}                 & Eq. 16&  \\
SxxXe131B& ${}^{131}Xe$ &\cite{Engel:1991wq}                  &
TABLE VII &  $S^{131}_{00}(0)$ should be $0.04$\footnote{As
confirmed by Bednyakov, private
communication}  \\
\hline
\end{tabular}
\end{table*}

\renewcommand{\theequation}{C-\arabic{equation}} 
  \setcounter{equation}{0}  
  \section*{Appendix C - Additional routines}  

To facilitate model independent comparisons with data, we 
provide additional routines to compute the nucleus recoil energy 
using as input the WIMP mass and  the cross sections for SI and SD scattering on nucleons.\\

\noindent
$\bullet$ \verb|nucleusRecoilAux(rho,fDv,A,Z,J,S00,S01,S11,Mwimp,csIp,csIn,csDp,csDn,dNdE)|\\
This function is similar to \verb|nucleusRecoil| and returns the number
of events per day and per $kg$ of detector material. The additional input parameters include
 the WIMP mass, \verb|csIp(csIn)| the SI cross sections for WIMP
scattering on protons (neutrons) and  \verb|csDp(csDn)| the SD cross sections for
protons(neutrons). A negative value for one of these cross sections corresponds to 
a destructive interference between proton and neutron amplitudes.

\noindent
$\bullet$ \verb|nucleusRecoil0Aux(rho,fDv,A,Z,J,Sp,Sn,Mwimp,csIp,csIn,csDp,csDn,dNdE)|\\
This function is similar  \verb|nucleusRecoilAux| except that it
uses Eq.~\ref{Sxx_0} to define $S00(0)$, $S11(0)$, $S01(0)$. For details see the description
of \verb|nucleusRecoil0|.

\noindent
$\bullet$ \verb|setRecoilEnergyGrid(step,nStep)|\\
This function  redefines the grid for the recoil energy. 
After calling this function the \verb|nucleusRecoil| function 
returns a recoil energy distribution where
$E_i= i\times {\rm step}$~keV,  $i=0,...,{\rm nStep}-1$.

\providecommand{\href}[2]{#2}\begingroup\raggedright\endgroup

\end{document}